\documentclass[conference]{IEEEtran}
\IEEEoverridecommandlockouts
\usepackage{cite}
\usepackage{tabularx}
\usepackage{braket}
\usepackage{amsmath,amssymb,amsfonts}
\usepackage{algorithmic}
\usepackage{graphicx}
\usepackage{textcomp}
\usepackage{xcolor}
\usepackage{subfigure}
\usepackage{subcaption}
\usepackage{soul} % Add this line to include the 'soul' package
\usepackage{caption} % Add the caption package
\usepackage{multirow}

\def\BibTeX{{\rm B\kern-.05em{\sc i\kern-.025em b}\kern-.08em
    T\kern-.1667em\lower.7ex\hbox{E}\kern-.125emX}}
\begin{document}

\title{Hardware Trojans in Quantum Circuits, Their Impacts, and Defense\\
% {\footnotesize \textsuperscript{*}Note: Sub-titles are not captured in Xplore and
% should not be used}
% \thanks{Identify applicable funding agency here. If none, delete this.}
}

\author{\IEEEauthorblockN{ Rupshali Roy}
\IEEEauthorblockA{\textit{School of EECS}\\
\textit{Penn State University,PA, USA}\\
rzr5509@psu.edu}
\and
\IEEEauthorblockN{Subrata Das}
\IEEEauthorblockA{\textit{School of EECS}\\
\textit{Penn State University,PA, USA}\\
sjd6366@psu.edu}
\and
\IEEEauthorblockN{Swaroop Ghosh}
\IEEEauthorblockA{\textit{School of EECS}\\ 
\textit{Penn State University,PA, USA}\\
szg212@psu.edu}
}

\maketitle

\begin{abstract}
The reliability of the outcome of a quantum circuit in near-term noisy quantum computers depends on the gate count and depth for a given problem. Circuits with a short depth and lower gate count can yield the correct solution more often than the variant with a higher gate count and depth. To work successfully for Noisy Intermediate Scale Quantum (NISQ) computers, quantum circuits need to be optimized efficiently using a compiler that decomposes high-level gates to native gates of the hardware. Many 3rd party compilers are being developed for lower compilation time, reduced circuit depth, and lower gate count for large quantum circuits. Such compilers, or even a specific release version of a compiler that is otherwise trustworthy, may be unreliable and give rise to security risks such as insertion of a quantum trojan during compilation that evades detection due to the lack of a golden/Oracle model in quantum computing. %Inherently, quantum circuits have a margin between the probability of correct and incorrect basis states. 
Trojans may corrupt the functionality to give flipped probabilities of basis states, or result in a lower probability of correct basis states in the output. In this paper, we investigate and discuss the impact of a single qubit Trojan (we have chosen a Hadamard gate and a NOT gate) inserted one at a time at various locations in benchmark quantum circuits without changing the the depth of the circuit. Results indicate an average of 16.18\% degradation for the Hadamard Trojan without noise, and 7.78\% with noise. For the NOT Trojan (with noise) there is 14.6\% degradation over all possible inputs. We then discuss the detection of such Trojans in a quantum circuit using CNN-based classifier achieving an accuracy of 90\%. %We conclude by discussing further possibilities of quantum Trojan insertion and detection.%The area of Quantum Computing (QC) has made great progress in recent times. Various qubit technologies have made the implementation of quantum algorithms on quantum computers practically feasible. In integrated circuits (ICs), hardware Trojan attacks have established themselves as an important security issue  [1]–[5]. As quantum circuits are becoming an important part of the computing landscape, it is essential to consider the Trojans that may infiltrate quantum computing systems as well. Quantum Trojans refer to circuit modifications  during design using an untrusted compiler. 
\end{abstract}


\section{Introduction}
Quantum computing is a fast-growing field with the potential to bring about a revolution in many industries, such as financial modeling, drug discovery and material science \cite{b1}, \cite{b2}. To perform computations, quantum computing uses the principles of quantum mechanics \cite{b3},\cite{b4} such as the ability to exist in multiple states at the same time. This unique property of qubits allows quantum computers to solve some complex problems exponentially faster than classical computers \cite{b4}. This is possible because qubits can maintain quantum coherence, i.e., they can stay in a superposition state without collapsing into a definite state. Various qubit technologies such as superconducting qubits \cite{b5}, trapped ion qubits \cite{b6}, photonic qubits \cite{b7}, quantum dots \cite{b8} and diamond nitrogen-vacancy centers \cite{b9}, have made the implementation of quantum algorithms on quantum computers practically feasible. The potential advantages offered by quantum computing have inspired significant research efforts globally. Tech giants like Microsoft, IBM and Google have already created prototype quantum computers \cite{b1}, \cite{b10}. In spite of the great possibilities in quantum computing, the technology is still in its infancy, and there are many challenges to overcome before it can be used widely \cite{b2}, \cite{b11}.

A major challenge in quantum computing is quantum circuit optimization. A quantum circuit is an ordered sequence of quantum gates that perform specific operations on qubits to solve a given problem \cite{b12}. Quantum circuits that have not been optimized well can produce random outputs instead of the desired results due to short coherence times and noise. Thus, optimization of quantum circuits has now become a prime research area in this field.it is a complex task requiring specialized tools and knowledge.Many quantum compilers have been developed, for example Qiskit, Forest and QuilC \cite{b13},\cite{b14}. They translate high-level descriptions of quantum circuits into low-level gates that can be executed on quantum hardware. The optimization quality and the time needed for compilation may vary significantly especially for complex quantum circuits. Untrusted third-party compilers have also come up, and they claim to provide faster and better optimization of large-scale quantum circuits than established compilers \cite{b15}, \cite{b16}.Using these, however, can result in severe security breaches, such as the introduction of quantum Trojans in the circuit during compilation. Such Trojans may either damage the functionality of the circuit, or cause quality degradation by reducing the probability of the correct basis state. This can result in considerable financial loss for the organization/researchers that originally developed the circuit and can also slow down the pace of the field of quantum computing. Thus, quantum circuits need to be optimized using novel techniques, while ensuring that they remain secure. In this paper we propose a single qubit gate based Trojan that can jeopardise the quantum circuit outputs. We also propose a machine learning based technique to detect such Trojans. 

Our study reveals several interesting aspects for tampering/Trojan insertion in quantum computing. For almost all the circuits that we tested, Trojans made of superposing gates like Hadamard gate cause the output quality to be degraded for the highest number of inputs when the Trojan is placed on one of the output qubits. The action of the Hadamard gate essentially causes the basis states to be superposed and this degrades the output quality.The following 2 points hold true for both kinds of Trojans we studied:(a) quantum circuits (especially the ones designed for arithmetic operations) have built in redundancy to tolerate errors and even functional corruption. This is primarily since basis state probability is used as output and there could be a significant margin between correct and incorrect basis state probabilities. Therefore, even one or two spurious gates may not corrupt the functionality, (b) some of the input vectors to the quantum circuit can be more tolerant to functional corruption than others. For example, 000 could tolerate few extra gates in the circuit and still give correct results than input 111 for a 3-input quantum circuit. This is true since the fake gates may not be sensitized for some inputs.     

\textbf{Paper Organization:} In Section II, we give background on quantum computing and review related works. This is followed by a discussion of the threat model, adversary capability and a description of quantum Trojans in Section III. Section IV presents the results. Section V describes the detection of Trojans using a CNN-based classifier and conclusions are drawn in Section VI.

%In the past decade, the area of Quantum Computing (QC) has made great progress. Thus QC now finds applications in fields such as machine learning  \cite{b6}, finance  \cite{b7}, chemistry  \cite{b8}, cybersecurity  \cite{b9}, and advanced manufacturing  \cite{b10}. Augmenting Quantum Random Access Memory (QRAM) can potentially provide exponential speedup for algorithms such as, Fourier transform  \cite{b11}, discrete logarithm  \cite{b12}, and pattern recognition  \cite{b13}– \cite{b15}.  
\section{Background}
We give a short introduction to a few principles fundamental to quantum computing that are relevant to our work.
\subsection{Quantum Computation Preliminaries} 
\subsubsection{Qubits} Analogous to classical bits, qubits are the smallest units of quantum information. A qubit can exist in a superposition of both states (0 and 1) simultaneously, as against classical bits that can exist in only one of the two states. This allows for multiple calculations to be carried out simultaneously. The most common way to represent a qubit is a two-level system, with the basis states $\ket{0}$ and $\ket{1}$. For a single qubit, the state can be represented as a linear combination of these two states, denoted by $\psi$ = $\alpha$ $\ket{0}$ + $\beta$ $\ket{1}$, where $\alpha$ and $\beta$ are complex numbers and the squared magnitudes of $\alpha$ and $\beta$ denote the probabilities of measuring the qubit in the states $\ket{0}$ and $\ket{1}$, respectively and the sum of these magnitudes is equal to 1\cite{b4}.

\subsubsection{Quantum Gates} Quantum gates are applied on qubits to perform specific operations such as entangling multiple qubits, changing the state of a qubit, and creating superposition states. Each quantum gate is represented by a unitary matrix which describes the transformation that is performed on the quantum state. Some commonly used quantum gates include the Hadamard gate, PauliX gate, and controlled-NOT (CNOT) gate. The Hadamard gate is used to create a superposition state of the $\ket{0}$ and $\ket{1}$ states with equal probability amplitudes. The CNOT gate is a two-qubit gate that applies the NOT operation to the target qubit if the control qubit is in the state $\ket{1}$, while the Pauli-X gate is used to flip the state of a qubit. A quantum gate is represented mathematically by a unitary matrix U, which satisfies the condition $\mathbf{U}^\intercal$U= I, where $\mathbf{U}^\intercal$\\ refers to the conjugate transpose of U and I is the identity matrix. The action of a quantum gate on a qubit state $\ket{\psi}$ is given by U$\ket{\psi}$. This action ensures the state vector is still normalised and denotes a rotation in the Bloch sphere representation of the qubit state.

\subsubsection{Coupling constraints and basis gates} A limited number of single and multi-qubit gates are supported by quantum computers in practice. These are known as basis gates or native gates of the hardware. IBM quantum computers, for example, use the following native gates: u1, u2, u3, id (single-qubit), and CNOT (two-qubit). However, the quantum circuit may contain high level gates that are not native to the hardware e.g., the Toffoli gate is not native to the IBM quantum computers. Therefore, the gates in a quantum circuit are decomposed into the basis gates before execution. Moreover, the two-qubit operation (CNOT) is allowed only between qubits that are connected. These restrictions in two-qubit operations in any target hardware are called coupling constraints.

\subsubsection{Compilation} Quantum circuit compilers e.g., Qiskit\cite{b17} carry out required steps (e.g., inserting SWAP gates) to the input circuits to meet coupling constraints of the hardware. Moreover, compilers provide higher-level circuit optimization using single/multi-qubit gate rotation, merging, cancellation and gate-reordering \cite{b18}. To restrict these additional optimizations across circuit partitions, Qiskit supports barriers between circuit partitions\cite{b17}.

\subsection{Untrusted Compiler-Oriented Attacks and Defenses}
%There have been many studies that have investigated the vulnerabilities associated with quantum computing. 
Some of the existing works have focused on the security risks associated with using unreliable quantum compilers \cite{b19},\cite{b20}. They point out the potential IP theft issues that may be introduced by untrustworthy compilers and propose obfuscation for defense. In another work \cite{b21}, the same risk is addressed by splitting the compilation process. Different sections of a quantum circuit are either sent to different compilers or to the same compiler at different times, thus giving only partial information to the adversary. This also gives factorial time reconstruction complexity.While other works have explored security risks to quantum computing from untrusted compilers, the relation among Trojans and their locations and arithmetic circuits and their inputs has not been investigated yet to the best of our knowledge. 
% In \cite{b22}, the authors study the effects Trojan insertion in QAOA circuits and detection using a machine learning model.
\section{Threat Model and Adversary Capability}
\subsection{Threat Model}
In this work, we consider the quantum circuits as valuable since it takes labor and time to create. 
We also assume that the user may employ unreliable/less-trusted third-party compilers. This is because of the scarcity of reliable compilers that provide state-of-the-art optimization results. However, using an untrusted compiler poses significant security risks, such as Trojan insertion by addition of quantum gates. The objectives could be to (a) corrupt the functionality or to (b) increase the computation time of the user in real hardware due to higher number of trials needed to estimate the basis state probabilities.

We assume that an unreliable third party hosts the compiler package remotely and a rogue adversary can take control of the compilation process to add undesirable gates to the quantum circuit in the process of optimization. Another scenario is that the compiler itself could be offered by an untrusted/rogue adversary. Although the adversary has access to the quantum circuit, he/she is unaware of the functionality. Therefore, the adversary will not be able to evaluate the impact of the Trojan insertion. Nevertheless, he/she can insert a suitable Trojan gate type in a suitable location based on prior experience with known quantum circuits. %location in the circuit that would cause the circuit functionality to be killed or degrade the quality of the output, %they may be able to tamper with the circuit such that the Trojan is not easily detectable, compromising the output of the computation and the integrity of the system, which is especially concerning when it is designed for use in critical situations such as financial analysis or national security.

\subsection{Adversary Capability}
We assume that the adversary (a) has access to the original or obfuscated quantum program. The users may choose to obfuscate the program using techniques from the literature \cite{b19, b21}, (b) has prior skills and computational resources to analyze the quantum program and identify Trojan gate type and the location and if inserted during compilation, can degrade the computation quality, (c) does not know the functionality or the correct outcome of the quantum circuit. This is reasonable assumption since on one hand, practical quantum programs cannot be simulated in classical computer while on the other, running quantum programs on real hardware will be expensive for the adversary.%can analyze the which may or may not be  by the 

\subsection{Quantum Trojan and Analysis Methodology}
%A quantum circuit consists of qubits, classical bits and quantum gates. Depending on the kind of functionality we wish to achieve, gates are chosen and placed in a pre-defined order on the qubits. For measurement of the output, we take the help of the classical bits. 
The adversary can insert Trojan gates in several ways. For example, they can insert a random gate/combination of gates from a collection of gate types or choose a specific gate/gate(s) and insert it into any part of the circuit, at the input and/or at the output or somewhere in the middle, during compilation.

%\subsection{Quantum Trojan Insertion Methodology}
The adversary will make the Trojan insertion as stealthy as possible. For example, if a circuit already has a certain kind of gate (say a CNOT gate) they will insert a Trojan of the same kind, since this would make it more difficult to detect its presence. Another way to make it tougher to detect the Trojan would be to ensure that the depth of the circuit remains unchanged after the insertion.

Our Trojan gates of choice are a single NOT gate and Hadamard gate. The NOT gate is classical i.e., it does not cause superposition while the Hadamard gate does. Hence we chose these gates to show the effects of superimposing and non-superimposing Trojans on quantum circuits. %since these gates are common in the quantum circuits.
They incur negligible overhead as compared to two-qubit gates. We placed the Trojan at all possible locations in the circuits such that the circuit depth remains same.
To ensure that all possible locations are covered, we iterate through each gate in the circuit in each layer, and identify the idle qubits i.e., qubits that are not involved in a gate operation. A NOT (or Hadamard) gate is placed on all such qubits one at a time. This process is repeated for all layers. Afterward, we get 2 variants of the original circuits each of which have a single NOT gate/single Hadamard gate inserted in a unique location. Fig. \ref{4gt13example} illustrates this for a 4gt13 circuit. 

We note that, there could be two types of Trojans, (a) one that does not change the probability distribution of the correct output basis state for a given input (i.e., probability of the correct basis state remains greater than 0.5), but instead degrades output quality, we call these non-flipping Trojans. For classical origin Trojans like NOT, this effect is seen only while simulating on noisy backends. But for superposing gates like Hadamard, this effect is seen even in the absence of noise. (b) Another that changes the probability distribution of the output basis states (i.e., brings down the probability of the correct state from more than 0.5 to less than 0.5) and are called flipping Trojans.

For better understanding, let us take two examples. 4gt11 circuit gives output 1 when the input is greater than 11.  Let it be given an input 0001. If the circuit were untainted, it would give output 0 with a probability higher than 50\%, 90.9 for this case where we ran the circuit for 1000 shots. This circuit is now attacked by a Trojan (NOT gate). The location of the Trojan gate is such that our circuit now perceives the input as a number which is still less than 11, although not 0001. Giving such an input to the 4gt11 circuit will still result in output 0 with a probability higher than 50\%, in this case 88\% (Fig. \ref{4gt11example}).

Along with the above-mentioned effects, there is also a third non-intuitive effect that is observed while applying the classical type NOT Trojan attacks on the circuits. For some Trojan locations, certain inputs may cause the circuit output to improve in quality, even when run for the same number of shots. For instance, if the output of the original circuit is 1 with a probability of 89\%, some Trojan attacks will result in the same output with a probability of 95\%. Moreover, this probability of basis states also changes every time one runs the circuit, whether infected or not. This means that the relative difference between performance of the attacked and original circuit may also vary. This occurs due to the spatial and temporal variation in noise for existing NISQ (noisy intermediate scale quantum) era quantum computers that are used for evaluation. While compiling quantum circuits in Qiskit, we use fake backends (FakeValencia in our case), which uses realistic noise models. The noise is modelled in a way that is akin to a random number generator. The amount of noise introduced during the simulation changes on every run, mimicking the temporal variation in noise that is observed in real quantum hardware. This randomly fluctuating noise sometimes improves/degrades the infected circuit output quality compared to the untainted circuit. However this random noise does not cause the functionality to be killed, it only affects the quality of the circuit output to vary over runs. 

Further, device noise is not uniform across all the qubits and the outcome distribution in such case may vary depending on the qubit mapping and hardware architecture which can affect SWAP overhead.Thus, the probability distribution of the circuit outputs will also vary over different backends %with different qubit mappings and SWAP overhead.
%While considering all these effects, we must also keep in mind that the quantum device noise is not uniformly distributed over all qubits. %\captionsetup[figure]{skip=2.5pt} % Adjust this value to control the spacing

\begin{figure} [t] 
    \centering
        \includegraphics[width=0.35\columnwidth]{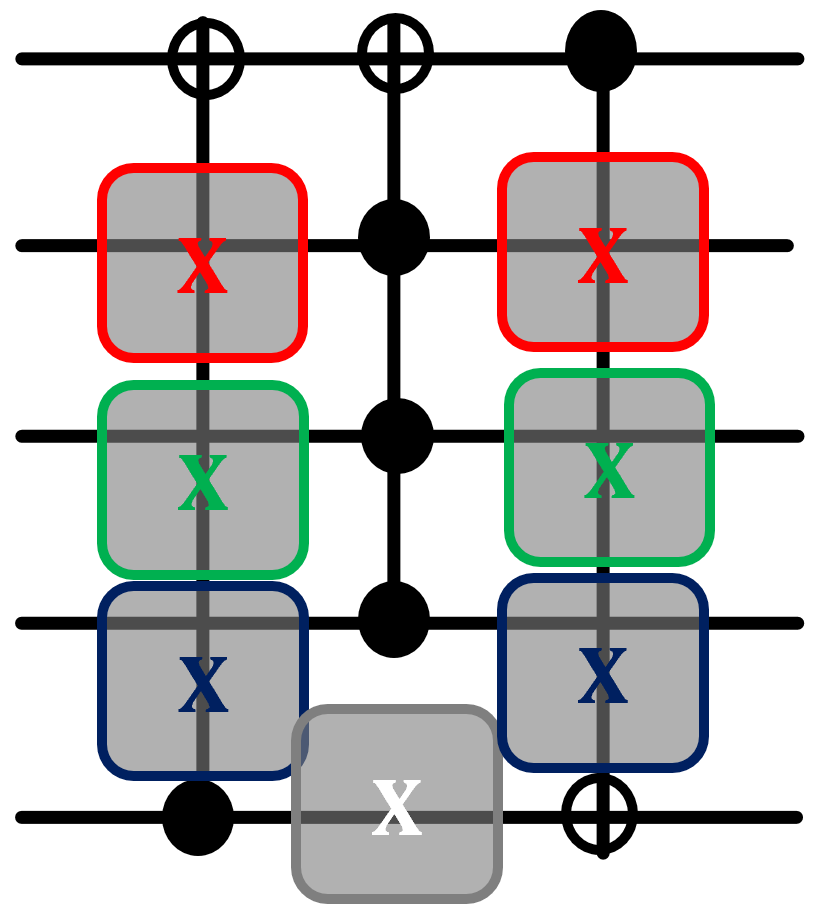}
         \vspace{-9mm}
        \label{4gt13example}
        \vspace{6mm}
        \caption{4gt13 benchmark circuit. Trojan gates located at idle locations for qubit 1(in red), for qubit 2(in green), for qubit 3 (in blue) and for qubit 4 (in white). 4 circuits are generated containing each of these 4 sets of Trojans, and then 7 circuits containing a single Trojan each are generated thereafter.}
        \label{4gt13example}         
\end{figure}

\section{Methodology and Results}
\subsection{Experimental Setup and Methods}
IBM Qiskit is used for the simulations run locally on an AMD Ryzen 5 5500U CPU with Radeon Graphics (2.10 GHz) machine with 8 GB RAM (Windows 11 Home). We selected 10 benchmark circuits from the Revlib \cite{b22} repository, which is commonly used for contemporary research on quantum circuit compilation. 
To investigate the effects of these Trojans, we run the original circuit and all variants of the infected circuits for 1000 shots, with all possible inputs. We use the FakeValencia backend of IBM Qiskit for simulating the Trojan affected circuits, which uses the realistic noise model of ibmq\_valencia device. For the Hadamard Trojan affected circuit, we also use the qasm\_simulator backend, which is an ideal backend without noise. We have chosen the backends in order to be able to showcase all the different effects of both kinds of Trojans in the absence as well as presence of noise.

\begin{figure} [t] 
    \centering
        \includegraphics[width=0.99\columnwidth]{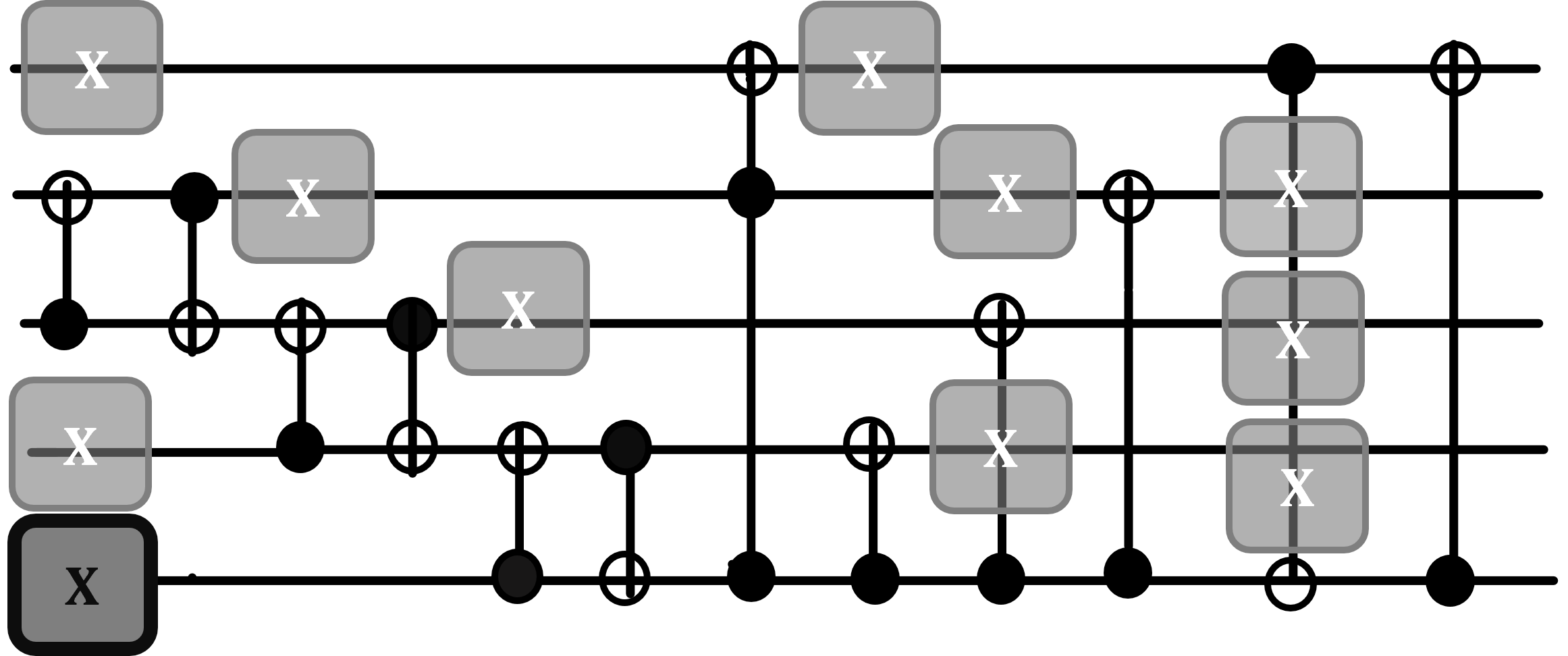}
         %\vspace{-9mm}
        \label{4gt11example}
        \vspace{-7mm}
        \caption{4gt11 benchmark circuit (in black) and all possible locations for single qubit Trojan gates(NOT in this case) shown (in gray). The specific Trojan gate being referred to is marked with a thick black border.}
        \label{4gt11example}  
\end{figure}
\subsection{Results}
\begin{figure*}[t] 
        \centering         
        \begin{minipage}{0.33\textwidth}
                \centering
                \includegraphics[width=\linewidth]{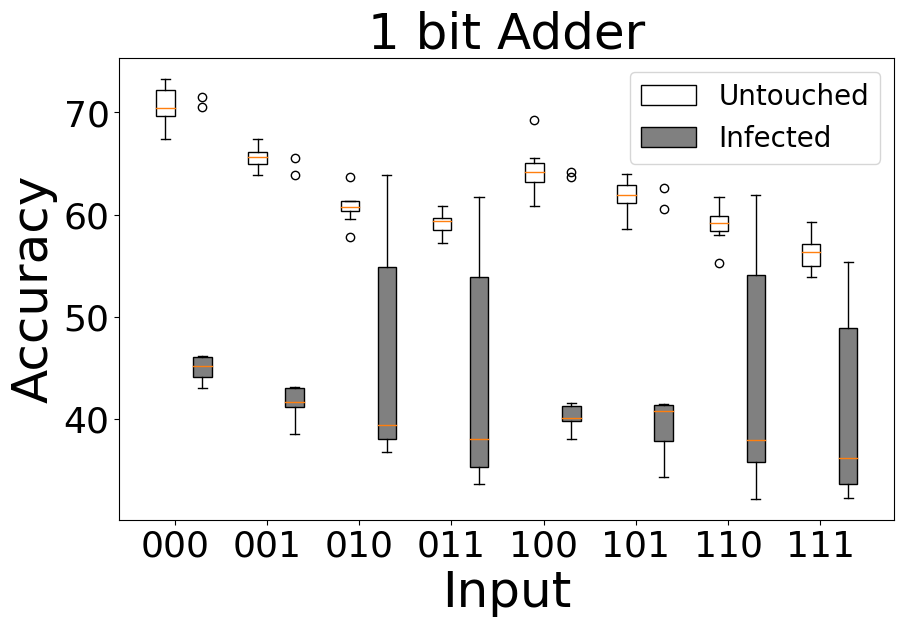}
                 \vspace{-5mm}
                 (a)
        \end{minipage}%  
        \begin{minipage}{0.33\textwidth}
                \centering
                \includegraphics[width=\linewidth]{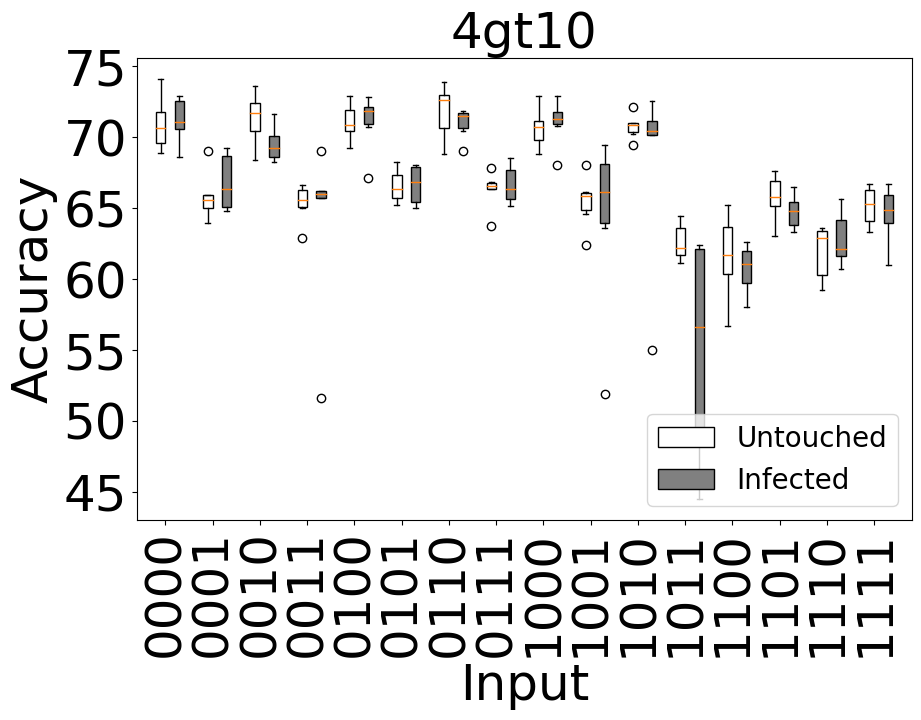}
                 \vspace{-5mm}
                 (b)
        \end{minipage}%
        \begin{minipage}{0.33\textwidth}
                \centering
                \includegraphics[width=\linewidth]{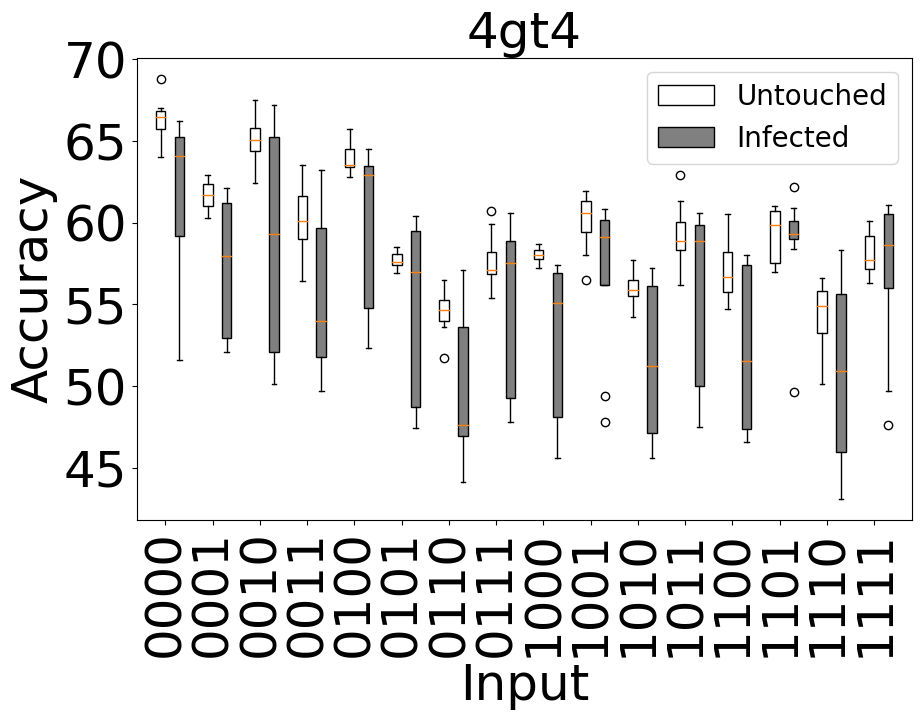}
                 \vspace{-5mm}
                 (c)
        \end{minipage}%
\end{figure*} 
\begin{figure*}
        \centering         
        \begin{minipage}{0.33\textwidth}
                \centering
                \includegraphics[width=\linewidth]{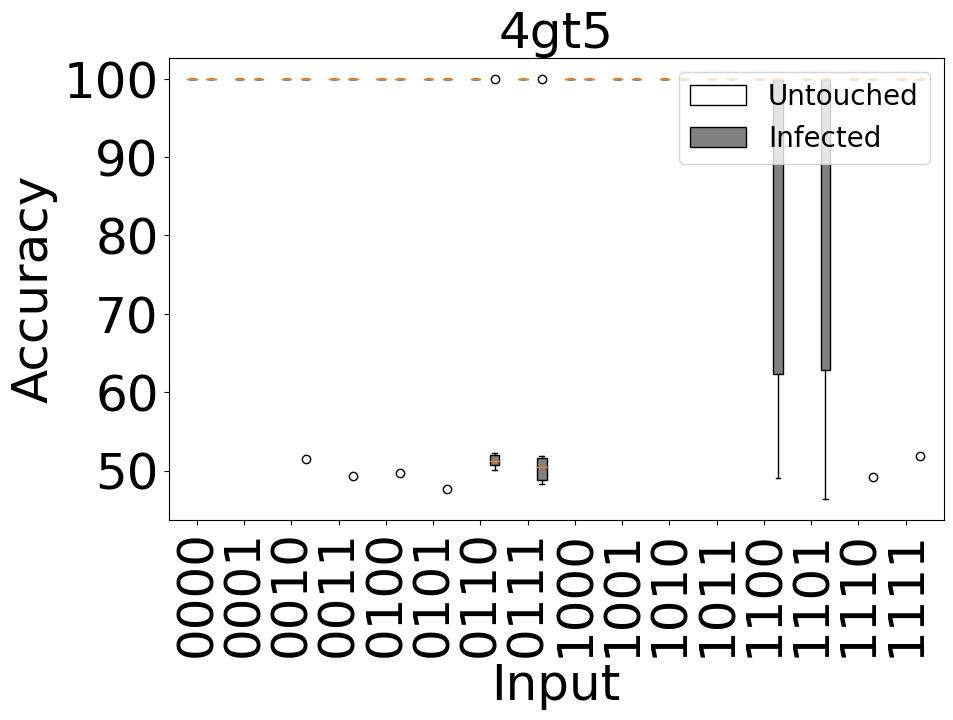}
                 \vspace{-5mm}
                 (d)
        \end{minipage}%  
        \begin{minipage}{0.33\textwidth}
                \centering
                \includegraphics[width=\linewidth]{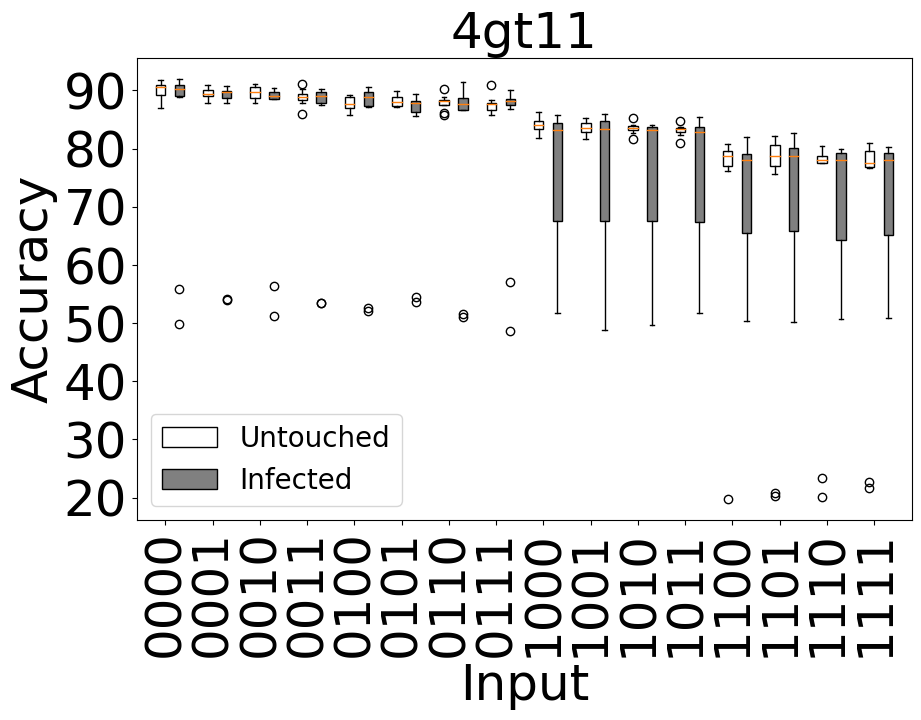}
                 \vspace{-5mm}
                 (e)
        \end{minipage}%
        \begin{minipage}{0.33\textwidth}
                \centering
                \includegraphics[width=\linewidth]{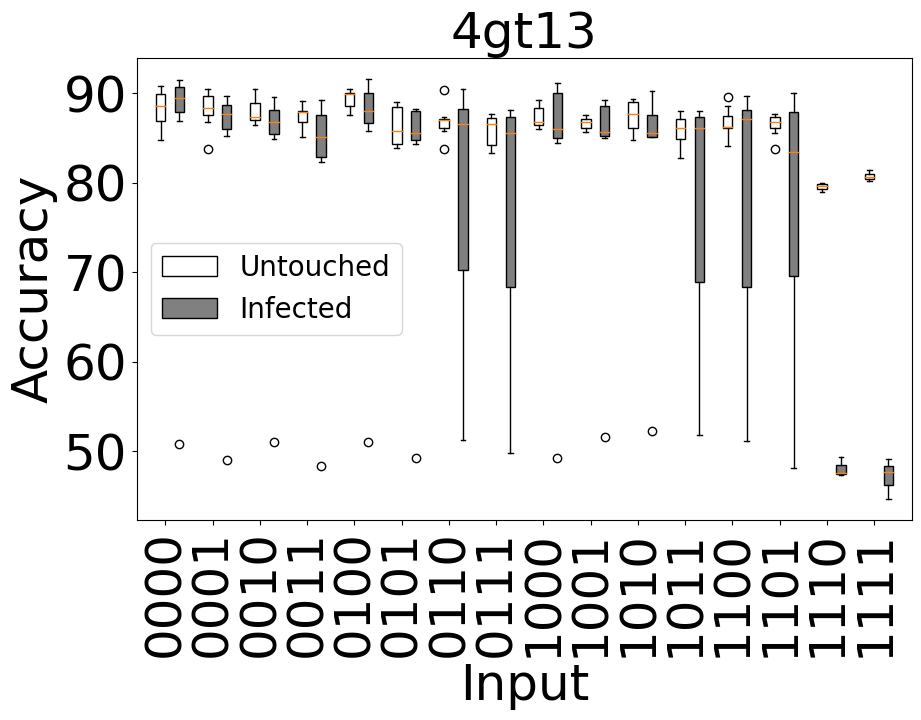}
                 \vspace{-5mm}
                 (f)
        \end{minipage}%
\end{figure*}
\begin{figure*} 
        \centering         
        \begin{minipage}{0.25\textwidth}
                \centering
                \includegraphics[width=\linewidth]{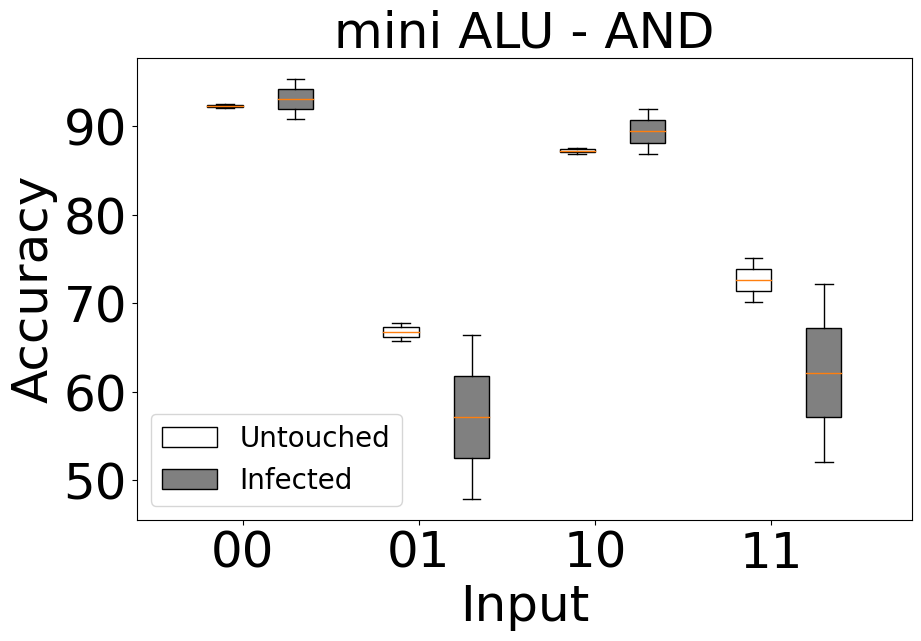}
                 \vspace{-5mm}
                 (g)
        \end{minipage}%  
        \begin{minipage}{0.25\textwidth}
                \centering
                \includegraphics[width=\linewidth]{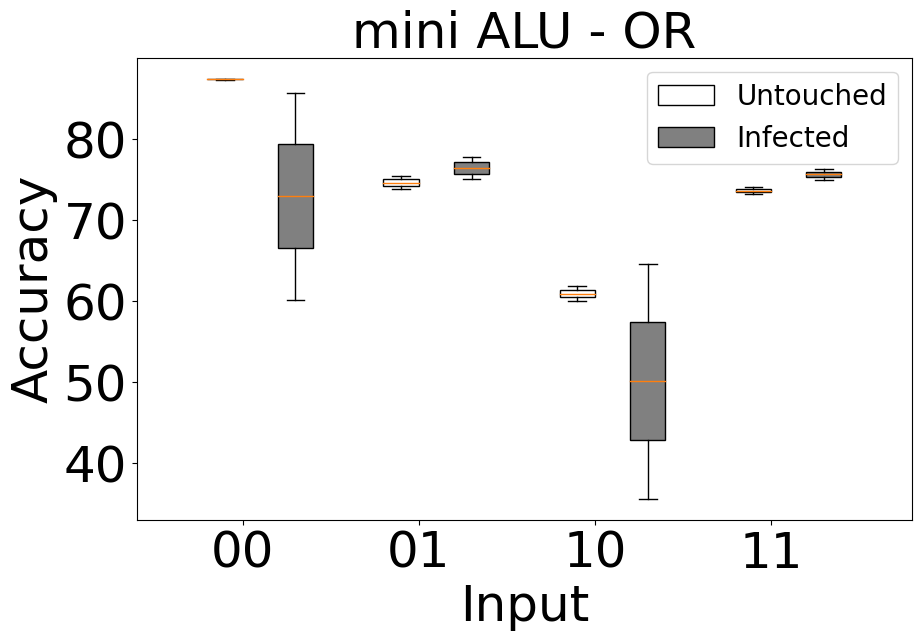}
                 \vspace{-5mm}
                 (h)
        \end{minipage}%
        \begin{minipage}{0.25\textwidth}
                \centering
                \includegraphics[width=\linewidth]{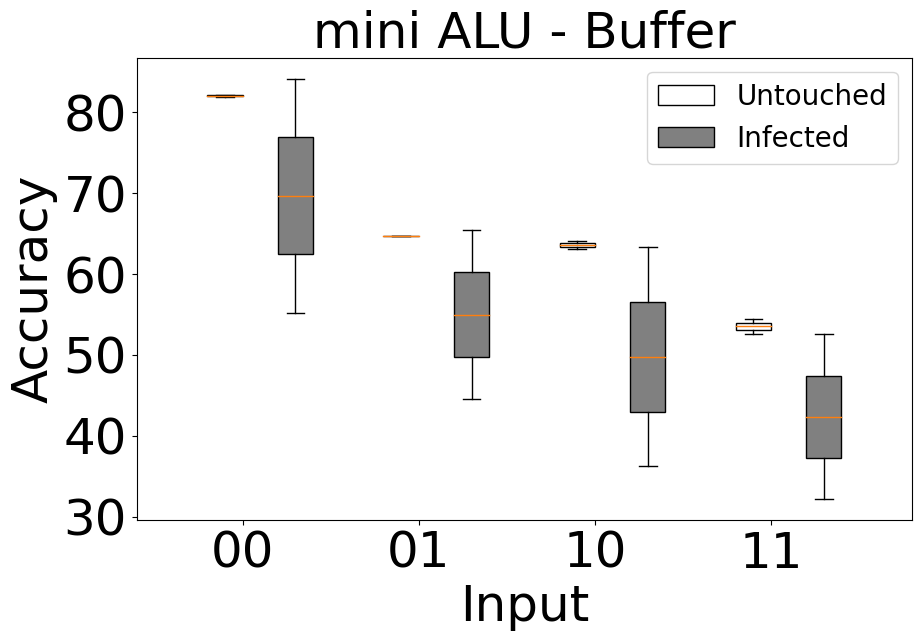}
                 \vspace{-5mm}
                (i)
        \end{minipage}%
        \begin{minipage}{0.25\textwidth}
                \centering
                \includegraphics[width=\linewidth]{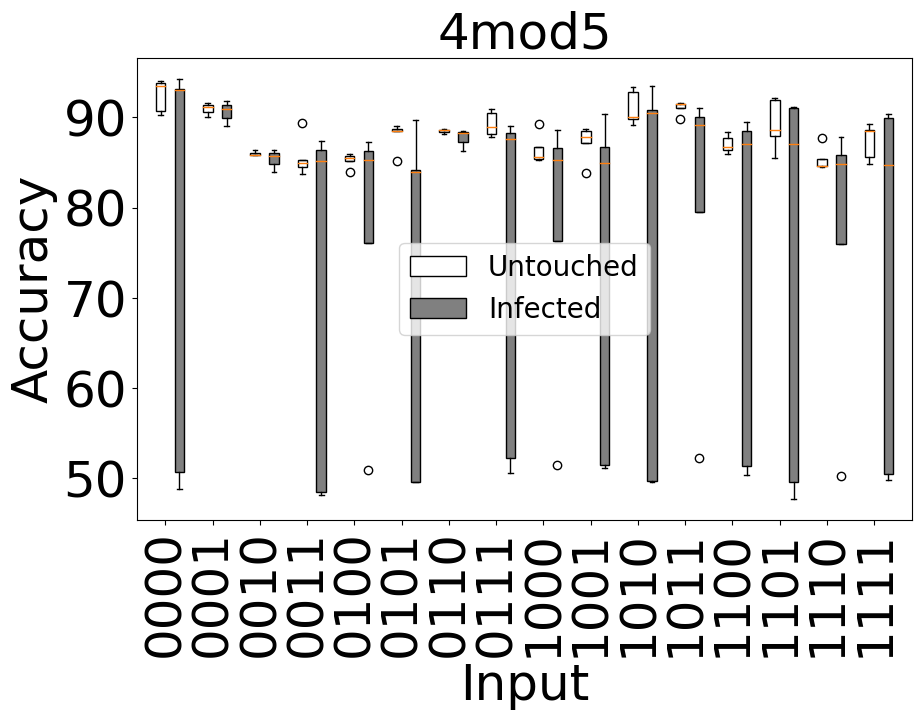}
                 \vspace{-4mm}
                 (j)
        \end{minipage}%
        \caption{Accuracy of Hadamard Trojan infected circuits in FakeValencia backend simulations with all possible inputs, as against untouched benchmark circuits, (a) 1 bit adder, (b) 4gt10, (c) 4gt4, (d) 4gt5, (e) 4gt11, (f) 4gt13, (g) Mini ALU - AND function, (h) Mini ALU - OR function, (i) Mini ALU - Buffer function and (j) 4mod5.} 
        \label{boxplots1}
\end{figure*} 

\begin{figure*}[t] 
        \centering         
        \begin{minipage}{0.33\textwidth}
                \centering
                \includegraphics[width=\linewidth]{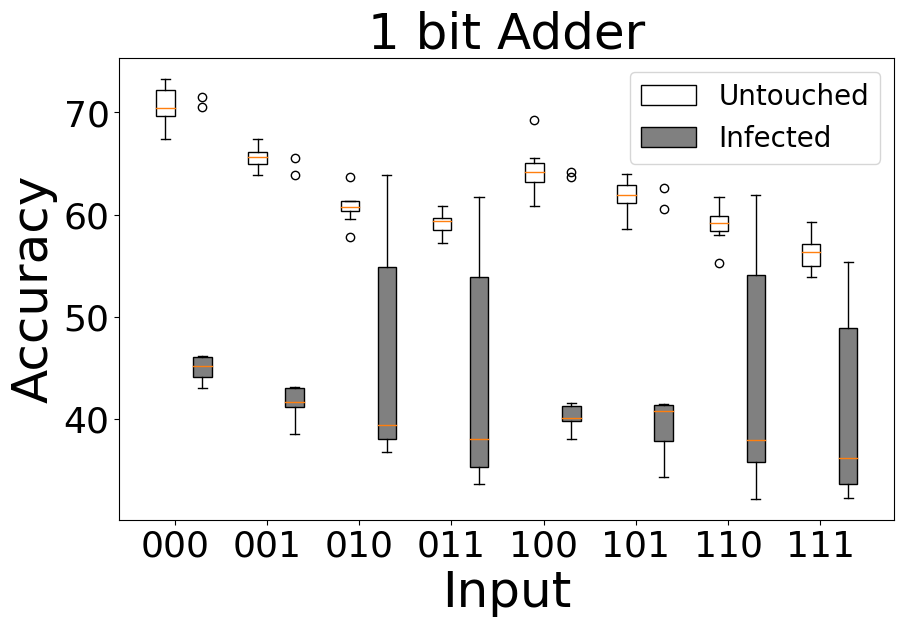}
                 \vspace{-6mm}
                 (a)
        \end{minipage}%  
        \begin{minipage}{0.33\textwidth}
                \centering
                \includegraphics[width=\linewidth]{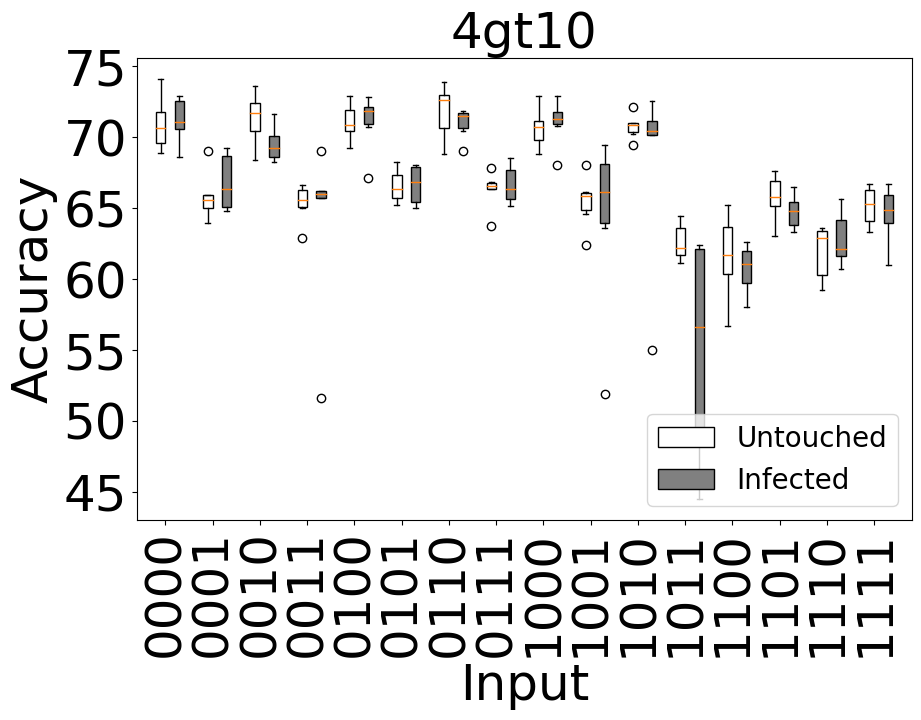}
                 \vspace{-6mm}
                 (b)
        \end{minipage}%
        \begin{minipage}{0.33\textwidth}
                \centering
                \includegraphics[width=\linewidth]{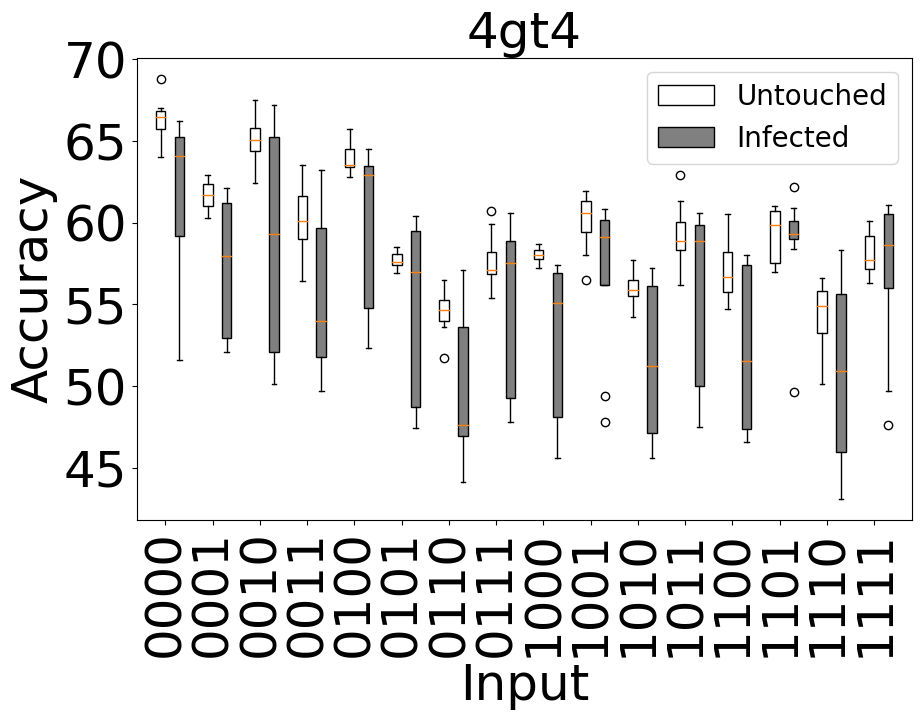}
                 \vspace{-6mm}
                 (c)
        \end{minipage}%
\end{figure*} 
\begin{figure*}
        \centering         
        \begin{minipage}{0.33\textwidth}
                \centering
                \includegraphics[width=\linewidth]{fig/4gt5_hplot.png}
                 \vspace{-6mm}
                 (d)
        \end{minipage}%  
        \begin{minipage}{0.33\textwidth}
                \centering
                \includegraphics[width=\linewidth]{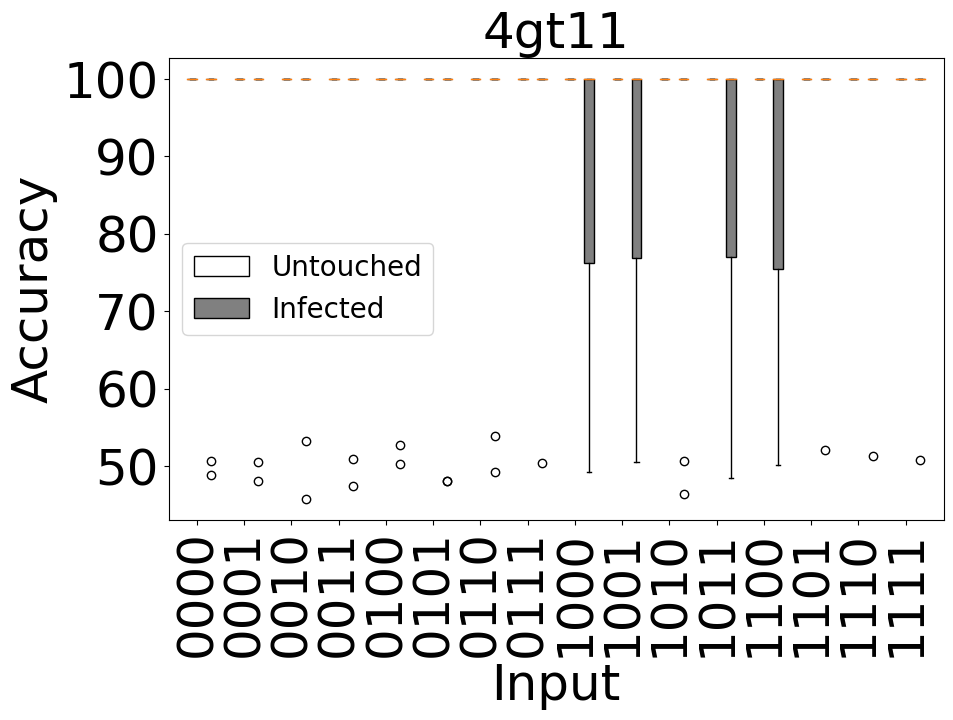}
                 \vspace{-6mm}
                 (e)
        \end{minipage}%
        \begin{minipage}{0.33\textwidth}
                \centering
                \includegraphics[width=\linewidth]{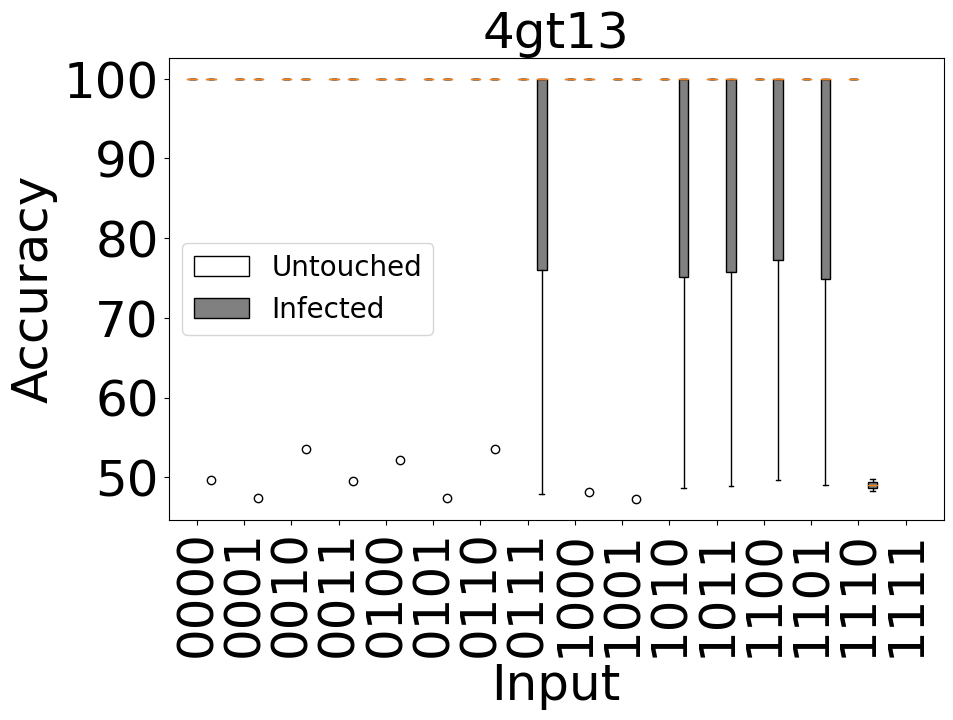}
                 \vspace{-6mm}
                 (f)
        \end{minipage}%
\end{figure*}
\begin{figure*} 
        \centering         
        \begin{minipage}{0.25\textwidth}
                \centering
                \includegraphics[width=\linewidth]{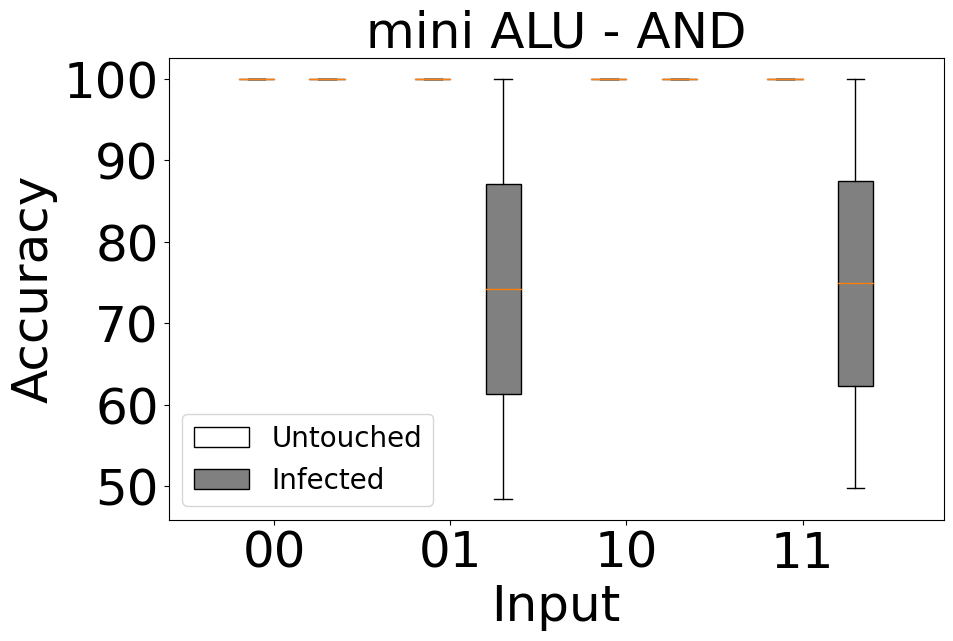}
                 \vspace{-5mm}
                 (g)
        \end{minipage}%  
        \begin{minipage}{0.25\textwidth}
                \centering
                \includegraphics[width=\linewidth]{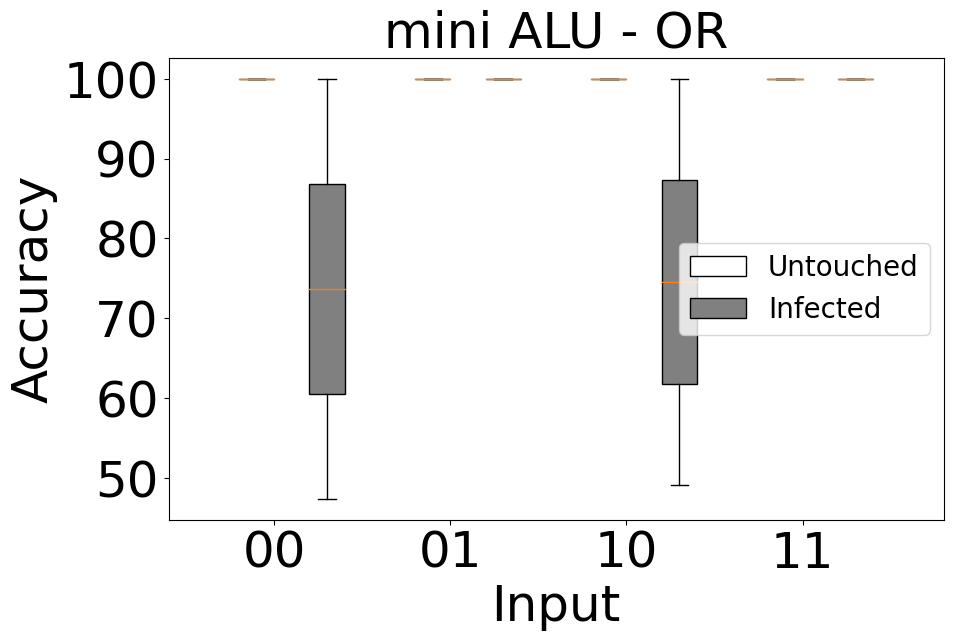}
                 \vspace{-5mm}
                 (h)
        \end{minipage}%
        \begin{minipage}{0.25\textwidth}
                \centering
                \includegraphics[width=\linewidth]{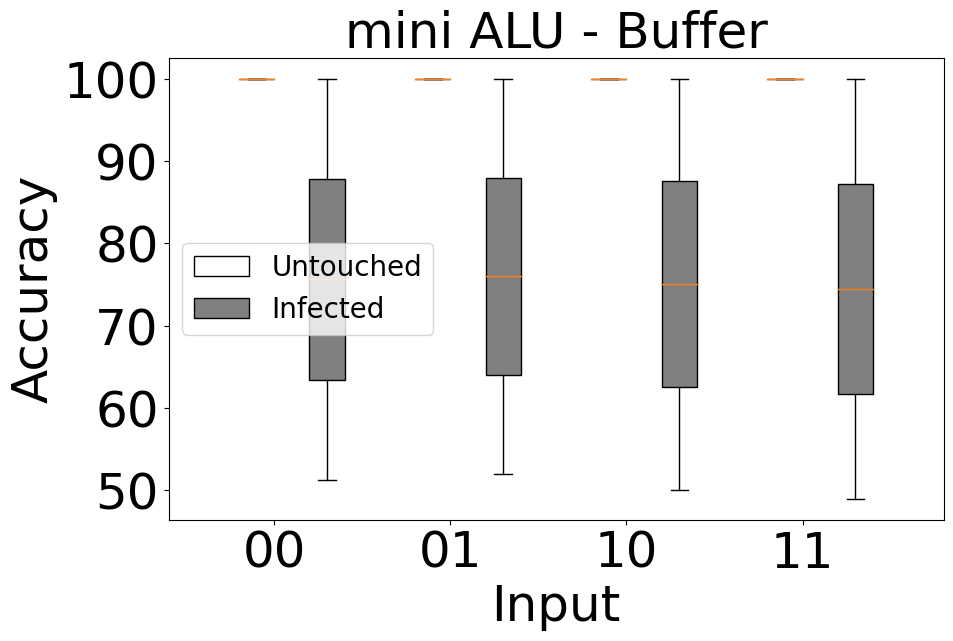}
                 \vspace{-5mm}
                (i)
        \end{minipage}%
        \begin{minipage}{0.25\textwidth}
                \centering
                \includegraphics[width=\linewidth]{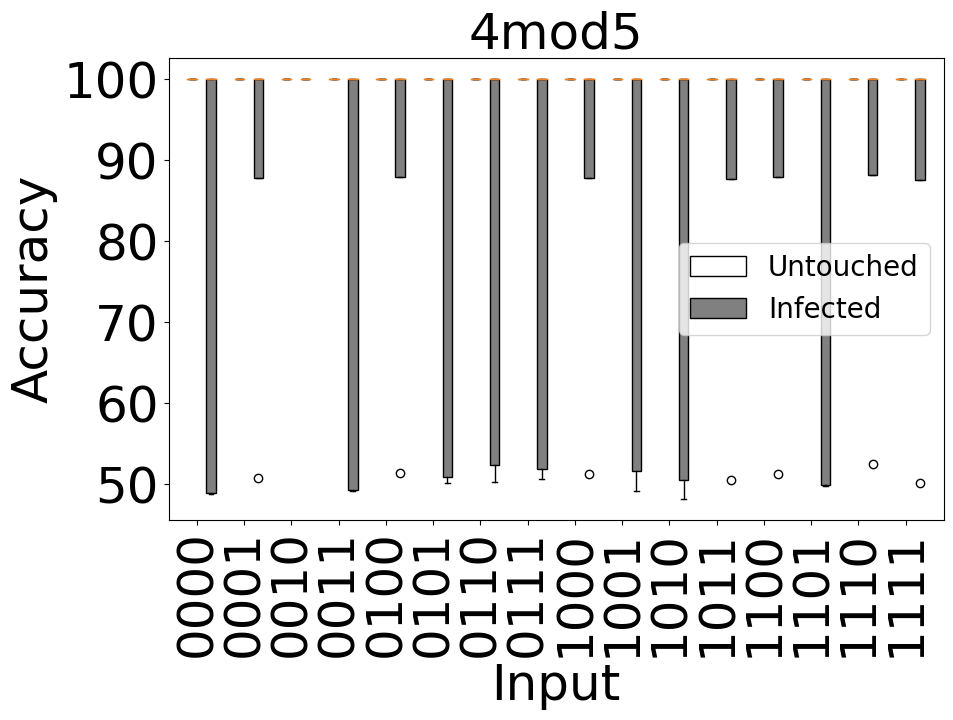}
                 \vspace{-5mm}
                 (j)
        \end{minipage}%
        \caption{Accuracy of Hadamard Trojan infected circuits in QASM simulator backend simulations with all possible inputs, as against untouched benchmark circuits, (a) 1 bit adder, (b) 4gt10, (c) 4gt4, (d) 4gt5, (e) 4gt11, (f) 4gt13, (g) Mini ALU - AND function, (h) Mini ALU - OR function, (i) Mini ALU - Buffer function and (j) 4mod5.} 
        \label{boxplots2}
\end{figure*}
\begin{figure*} 
        \centering         
        \begin{minipage}{0.25\textwidth}
                \centering
                \includegraphics[width=\linewidth]{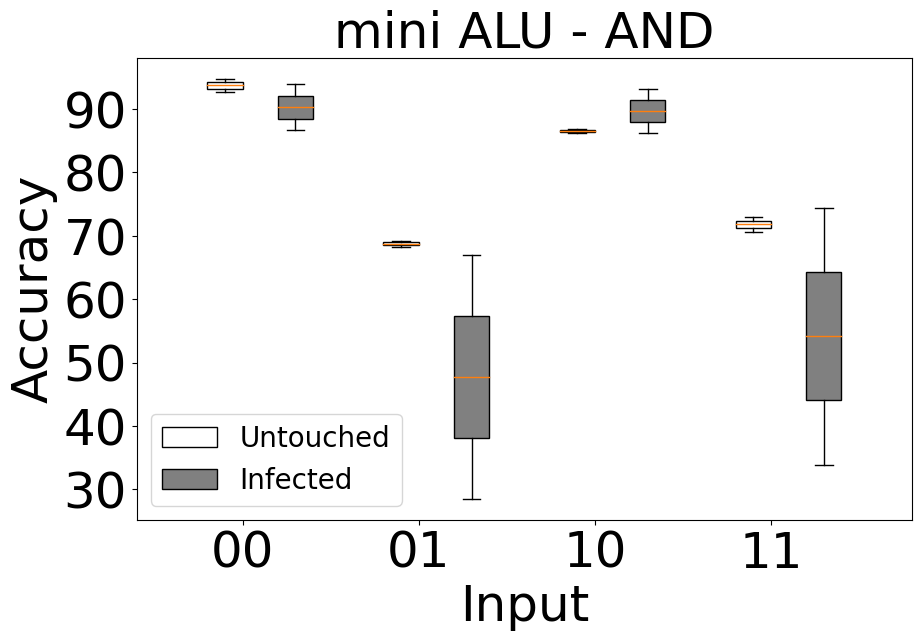}
                 \vspace{-4mm}
                 (g)
        \end{minipage}%  
        \begin{minipage}{0.25\textwidth}
                \centering
                \includegraphics[width=\linewidth]{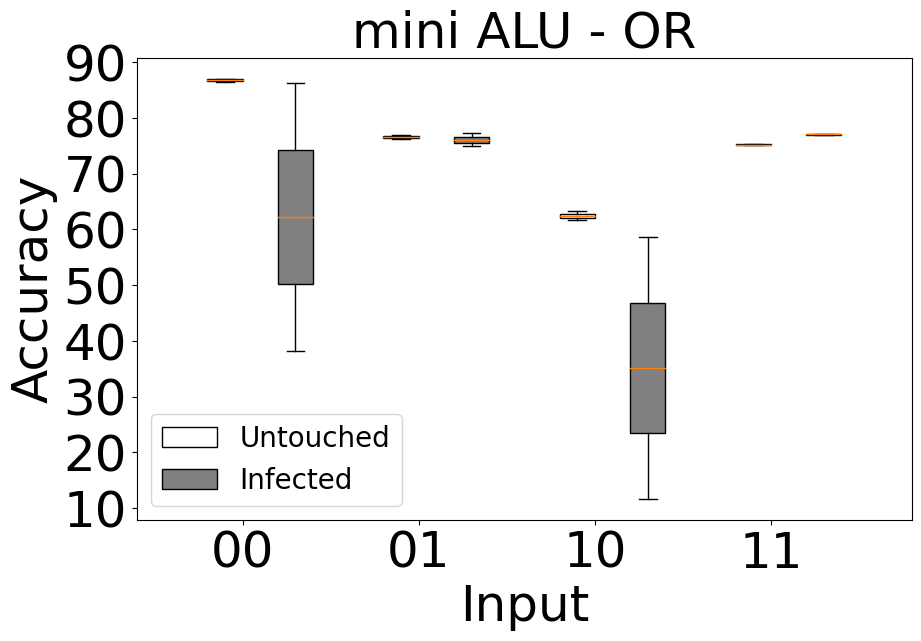}
                 \vspace{-4mm}
                 (h)
        \end{minipage}%
        \begin{minipage}{0.25\textwidth}
                \centering
                \includegraphics[width=\linewidth]{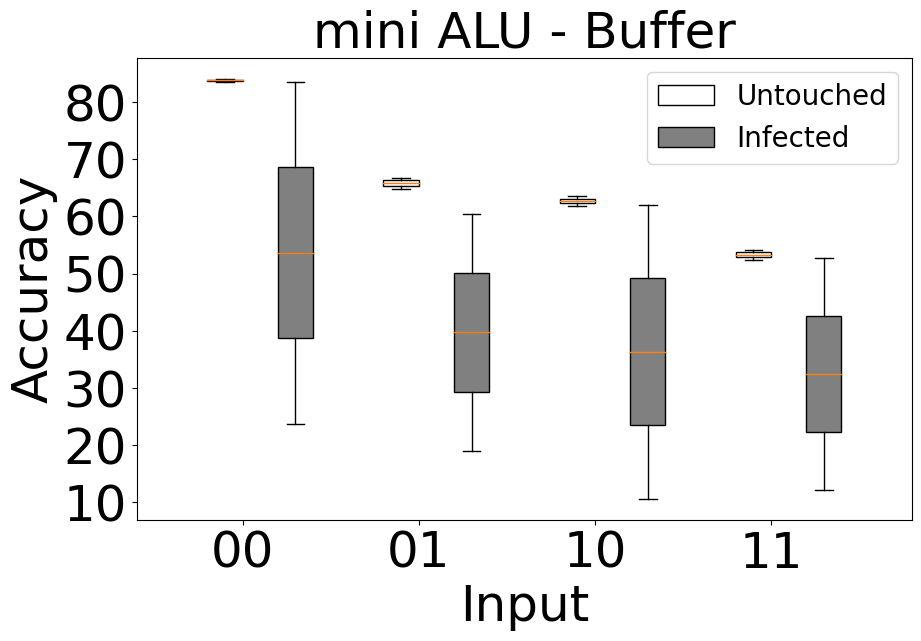}
                 \vspace{-4mm}
                (i)
        \end{minipage}%
        \begin{minipage}{0.25\textwidth}
                \centering
                \includegraphics[width=\linewidth]{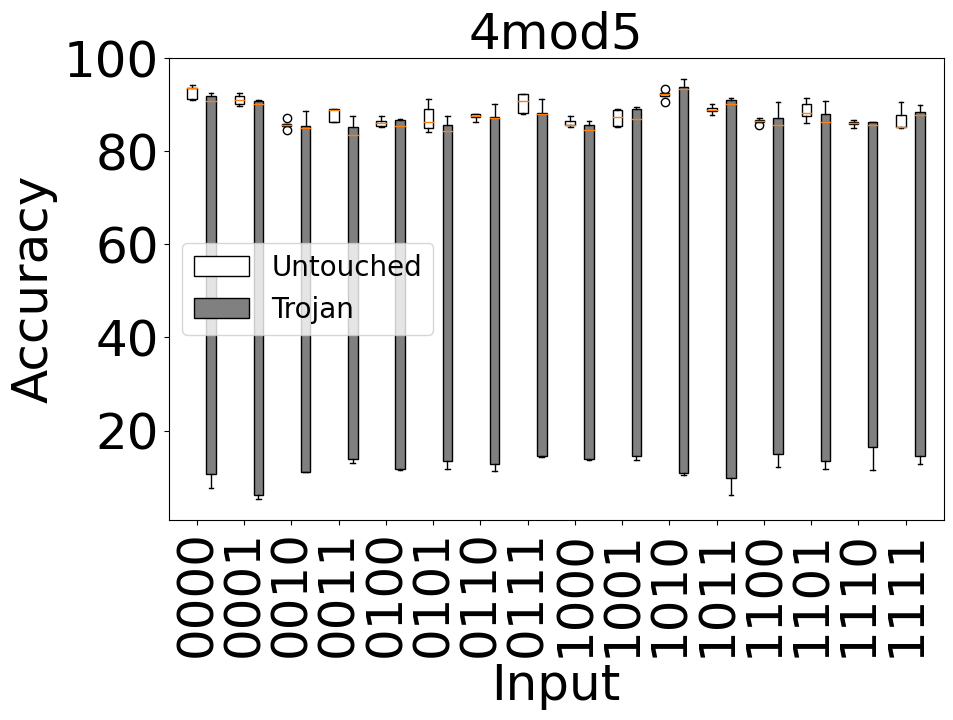}
                 \vspace{-4mm}
                 (j)
        \end{minipage}%
        \caption{Accuracy of NOT type Trojan infected circuits with all possible inputs, as against untouched benchmark circuits, (a) 1 bit adder, (b) 4gt10, (c) 4gt4, (d) 4gt5, (e) 4gt11, (f) 4gt13, (g) Mini ALU - AND function, (h) Mini ALU - OR function, (i) Mini ALU - Buffer function and (j) 4mod5.} 
        \label{boxplots3}
\end{figure*}

  
To present the results of our investigation, we plot the accuracies of the outputs for all the Trojan-infected circuits for a given benchmark against the associated input values, and also plot the same for untouched benchmarks (Fig. \ref{boxplots1}, \ref{boxplots2} and \ref{boxplots3}). Accuracy in this context, refers to the probability of the correct basis state in the output of a quantum circuit. 

In these plots we observe that the accuracies for the untouched benchmarks are high and show very little variation, as is expected. The random noise modelled into FakeValencia gives rise to the small variations around the median for those simulations. For different Trojans and different circuit inputs, we see following effects on the accuracy.

1) For some inputs, we observe that the accuracies are very low and do not vary a lot either. This means that for the said input, almost all the Trojans are flipping, i.e., the functionality of the circuit has been corrupted.

2) Some inputs give rise to accuracies that vary over a large range; some very low, while others are high. This implies that some Trojans are flipping, while some are non-flipping for this input. The position of the median line tells us which kind of Trojans dominate for the input in question. If the line is closer to the minimum accuracy, this means that the flipping Trojans are in the majority, and vice versa. 

3) For some inputs, the accuracies are high, and vary little. The median accuracy, in most cases, will be lower than that for the untouched benchmarks. Most of the Trojans in this case are non-flipping. For the FakeValencia backend, due to the random noise modeled into the simulator used, a few median accuracies even appear to go higher than the median accuracy for the corresponding untouched circuit. 

From these results, we also gain some insights into how some inputs are more resilient than others. For example, in the 4gt(x) series when affected by the NOT Trojan type, we note that most inputs less than x are less susceptible to degradation - this is especially evident in Fig. \ref{boxplots2}(b), (d) and (e). For mini ALU - AND function we observe that odd inputs are more susceptible to degradation, while for the OR functionality it is the other way round, as noted in Fig. \ref{boxplots2}(g) and (h). For other circuits such as 4gt4, mini ALU - buffer and 4mod5, shown in Fig. \ref{boxplots2}(c),(i) and (j), all inputs are equally vulnerable to degradation by Trojans.

On an average over all inputs and benchmarks in our simulation for the NOT Trojan type, we note a maximum degradation of 85.2\% for flipping Trojan types. For non-flipping Trojans, the maximum degradation observed is 13.4\%. The aggregate average degradation over all inputs for all Trojan locations over all benchmarks is found to be 14.6\%. 
For the Hadamard Trojan simulations in an ideal QASM simulator, we note a maximum degradation of 54.7\% and 45.2\% on FakeValencia for flipping Trojan types on an average over all inputs and benchmarks in our simulation. For non-flipping Trojans, the maximum degradation observed is 49.9\% in a QASM simulator and 40.4\% in FakeValencia . The aggregate average degradation over all inputs for all Trojan locations over all benchmarks is found to be 16.18\% in QASM simulator and 7.78\% in FakeValencia.In the following table, we see a summary of the degradation in outputs of Trojan infected circuits. The minus sign in the minimum degradation values for non-flipping Trojans for Fakevalencia simulations signifies that in these cases, the infected circuit yielded better output probability for the correct basis state. This can be attributed to the dynamic variation in noise in FakeValencia and other backends.  

%\small
\begin{figure*}
\centering
\caption{Degradation of outputs of Trojan infected circuits}
\label{1}
\begin{tabular}{cll|ll|l|}
\cline{4-6}
\multicolumn{1}{l}{}                                   &                                                    &         & \multicolumn{2}{c|}{Hadamard Trojan}               & NOT Trojan   \\ \cline{4-6} 
\multicolumn{1}{l}{}                                   &                                                    &         & \multicolumn{1}{l|}{QASM simulator} & FakeValencia & FakeValencia \\ \hline
\multicolumn{1}{|c|}{\multirow{7}{*}{Degradation(\%)}} & \multicolumn{1}{l|}{\multirow{3}{*}{Flipping}}     & Maximum & \multicolumn{1}{l|}{54.7}           & 45.2         & 85.2         \\ \cline{3-6} 
\multicolumn{1}{|c|}{}                                 & \multicolumn{1}{l|}{}                              & Minimum & \multicolumn{1}{l|}{50.1}           & 3.2          & 13.4         \\ \cline{3-6} 
\multicolumn{1}{|c|}{}                                 & \multicolumn{1}{l|}{}                              & Overall & \multicolumn{1}{l|}{51.46}          & 25.87        & 51.2         \\ \cline{2-6} 
\multicolumn{1}{|c|}{}                                 & \multicolumn{1}{l|}{\multirow{3}{*}{Non Flipping}} & Maximum & \multicolumn{1}{l|}{49.9}           & 40.4         & 13.4         \\ \cline{3-6} 
\multicolumn{1}{|c|}{}                                 & \multicolumn{1}{l|}{}                              & Minimum & \multicolumn{1}{l|}{0}              & -6.4         & -8.2         \\ \cline{3-6} 
\multicolumn{1}{|c|}{}                                 & \multicolumn{1}{l|}{}                              & Overall & \multicolumn{1}{l|}{8.83}           & 3.21         & 0.11         \\ \cline{2-6} 
\multicolumn{1}{|c|}{}                                 & \multicolumn{1}{l|}{Overall}                       &         & \multicolumn{1}{l|}{16.18}          & 7.78         & 14.6         \\ \hline
\end{tabular}
\end{figure*}
Based on the results presented above, some locations and inputs may be more favorable for the adversary to corrupt the functionality. However, the adversary does not have control over the inputs. With respect to location of Trojan  The trend with respect to the location of Trojan insertion is not well-defined for NOT gates. Nevertheless, adding NOT gates at the measured qubits close to the output may increase the chances of functional corruption. Under this scenario, we assume that the adversary may insert a Trojan at any of the locations randomly or focus on measured qubits closer to the output.
        % \begin{subfigure}{}
        %         \centering
        %         \includegraphics[width=0.45\columnwidth]{fig/4gt5.png}
        %          \vspace{-0mm}
        %         \label{4gt5}
        % \end{subfigure}
        % \begin{subfigure}{}
        %         \centering
        %         \includegraphics[width=0.45\columnwidth]{fig/4gt11.png}
        %          \vspace{-0mm}
        %         \label{4gt11}
        % \end{subfigure}
        % \begin{subfigure}{}
        %         \centering
        %         \includegraphics[width=0.45\columnwidth]{fig/4gt13.png}
        %          \vspace{-0mm}
        %         \label{4gt13}
        % \end{subfigure}
        % \begin{subfigure}{}
        %         \centering
        %         \includegraphics[width=0.43\columnwidth]{fig/4mod5.png}
        %          \vspace{-0mm}
        %         \label{4mod5}
        % \end{subfigure}
        % \begin{subfigure}{}
        %         \centering
        %         \includegraphics[width=0.45\columnwidth]{fig/AND.png}
        %          \vspace{-0mm}
        %         \label{and}
        % \end{subfigure}
        % \begin{subfigure}{}
        %         \centering
        %         \includegraphics[width=0.45\columnwidth]{fig/or.png}
        %          \vspace{-0mm}

        %         \label{or}
        % \end{subfigure}
        % \begin{subfigure}{}
        %         \centering
        %         \includegraphics[width=0.45\columnwidth]{fig/buffer.png}
        %          \vspace{-0mm}
        %         \label{buffer}
        % \end{subfigure}
\begin{figure} [h] 
    \begin{subfigure}{}
        \centering
        \includegraphics[width=0.45\columnwidth]{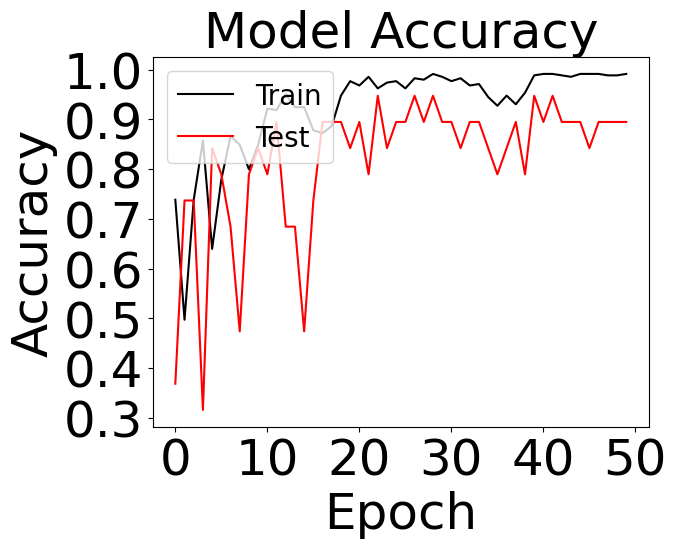}
        \vspace{-6mm}
        \label{accu}
        \vspace{3mm}
    \end{subfigure}%
    \begin{subfigure}{}
        \centering
        \includegraphics[width=0.45\columnwidth]{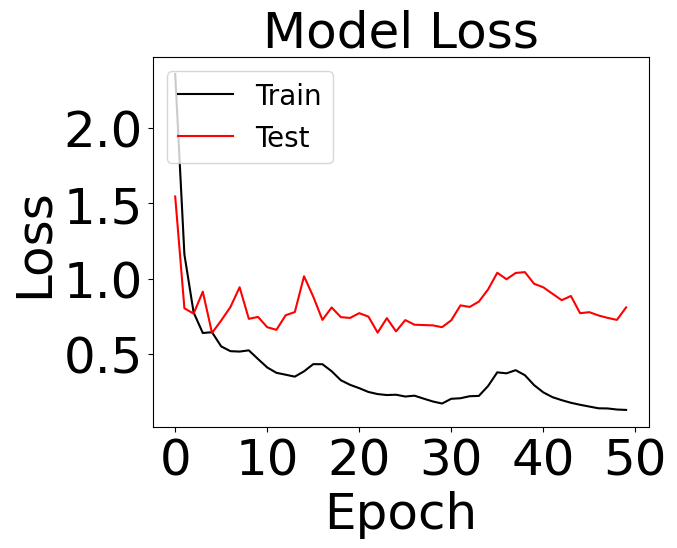}
        \vspace{-5mm}
        \label{loss}
        \vspace{3mm}
    \end{subfigure}%
    \caption{Accuracy and loss versus epochs plots demonstrating the classifier's training progress for the NOT infected circuit dataset.}
    \label{accu_vs_epoch_not}
\end{figure}

\begin{figure} [h] 
    \begin{subfigure}{}
        \centering
        \includegraphics[width=0.45\columnwidth]{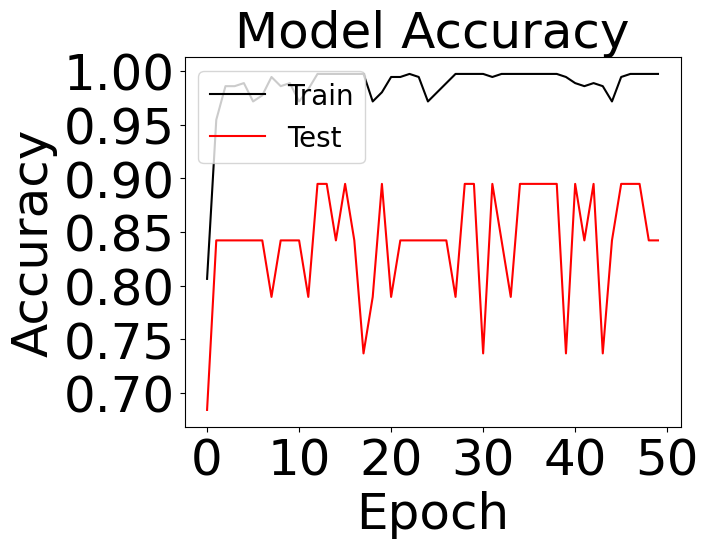}
        \vspace{-6mm}
        \label{accu}
        \vspace{3mm}
    \end{subfigure}%
    \begin{subfigure}{}
        \centering
        \includegraphics[width=0.45\columnwidth]{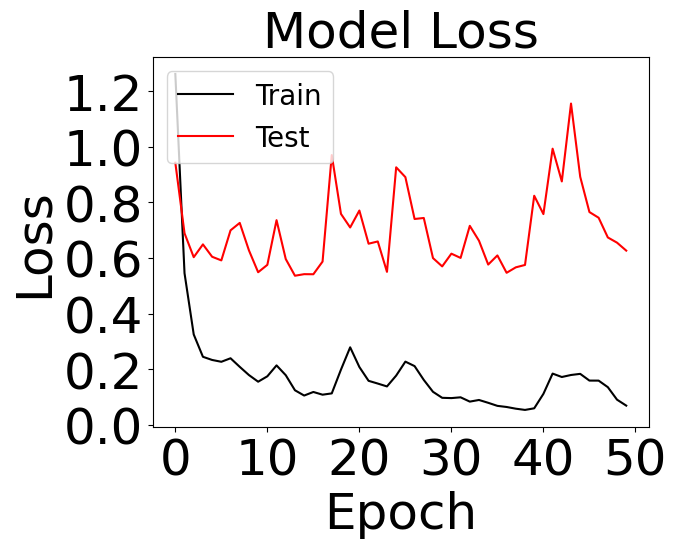}
        \vspace{-5mm}
        \label{loss}
        \vspace{3mm}
    \end{subfigure}%
    \caption{Accuracy and loss versus epochs plots demonstrating the classifier's training progress for the Hadamard infected circuit dataset.}
    \label{accu_vs_epoch_h}
\end{figure}
\vspace{-3mm}

\begin{figure*}[t] 
        \centering         
        \begin{minipage}{0.33\textwidth}
                \centering
                \includegraphics[width=\linewidth]{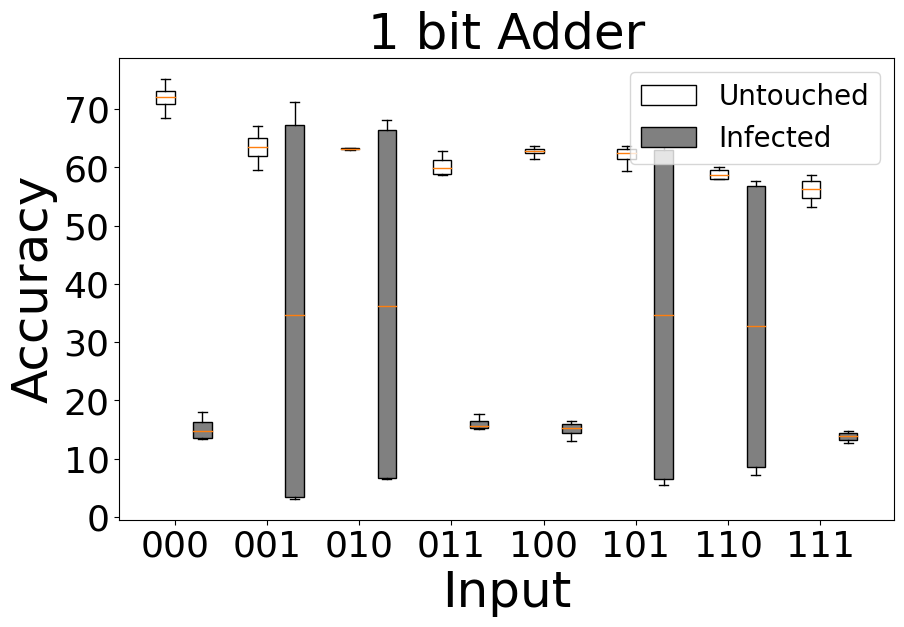}
                 \vspace{-4mm}
                 (a)
        \end{minipage}%  
        \begin{minipage}{0.33\textwidth}
                \centering
                \includegraphics[width=\linewidth]{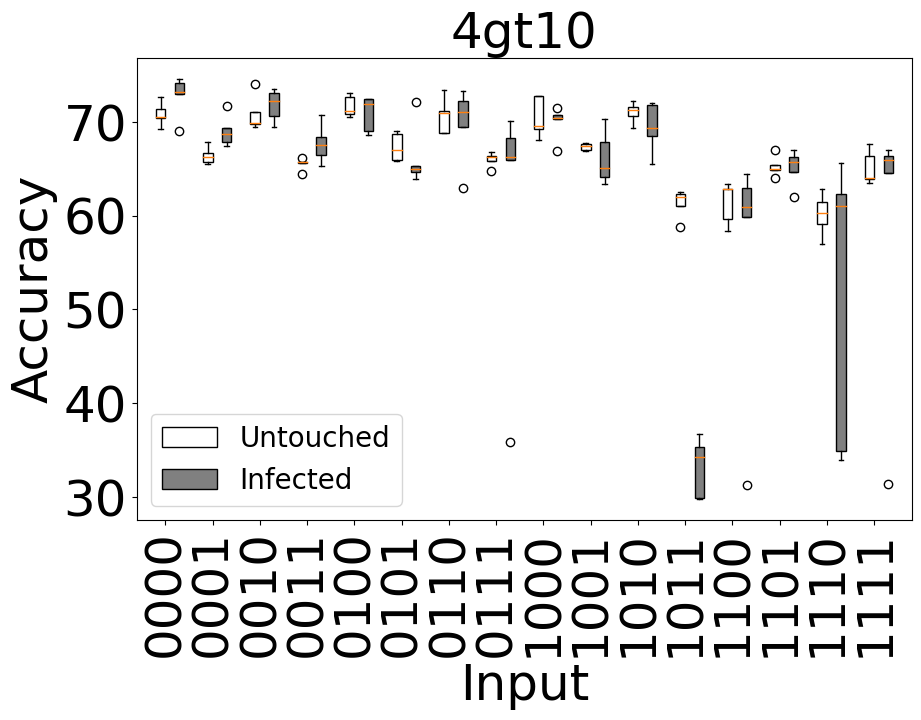}
                 \vspace{-4mm}
                 (b)
        \end{minipage}%
        \begin{minipage}{0.33\textwidth}
                \centering
                \includegraphics[width=\linewidth]{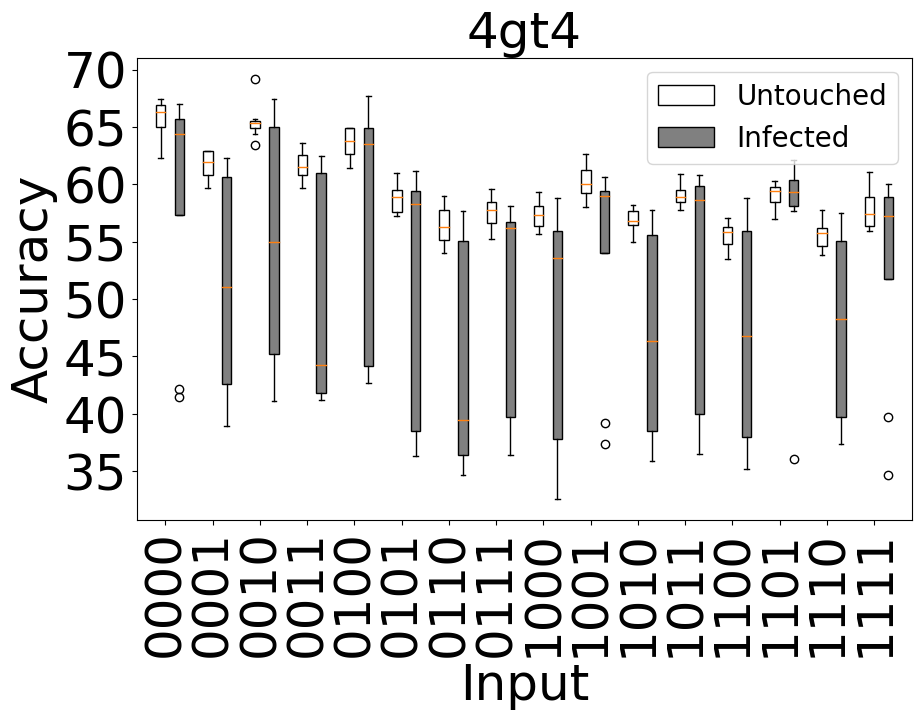}
                 \vspace{-4mm}
                 (c)
        \end{minipage}%
\end{figure*} 
\begin{figure*}
        \centering         
        \begin{minipage}{0.33\textwidth}
                \centering
                \includegraphics[width=\linewidth]{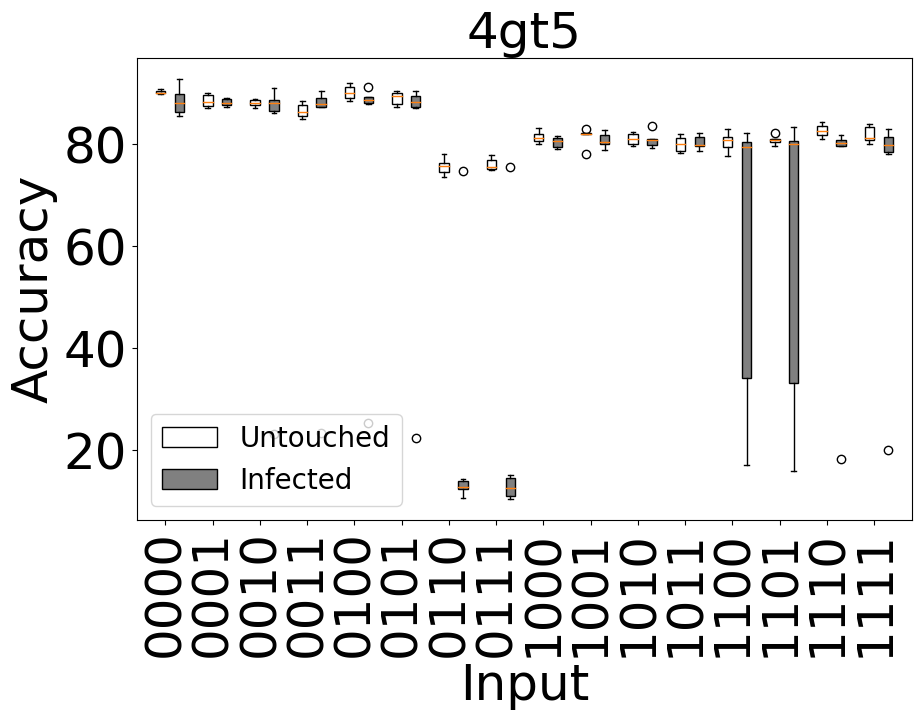}
                 \vspace{-4mm}
                 (d)
        \end{minipage}%  
        \begin{minipage}{0.33\textwidth}
                \centering
                \includegraphics[width=\linewidth]{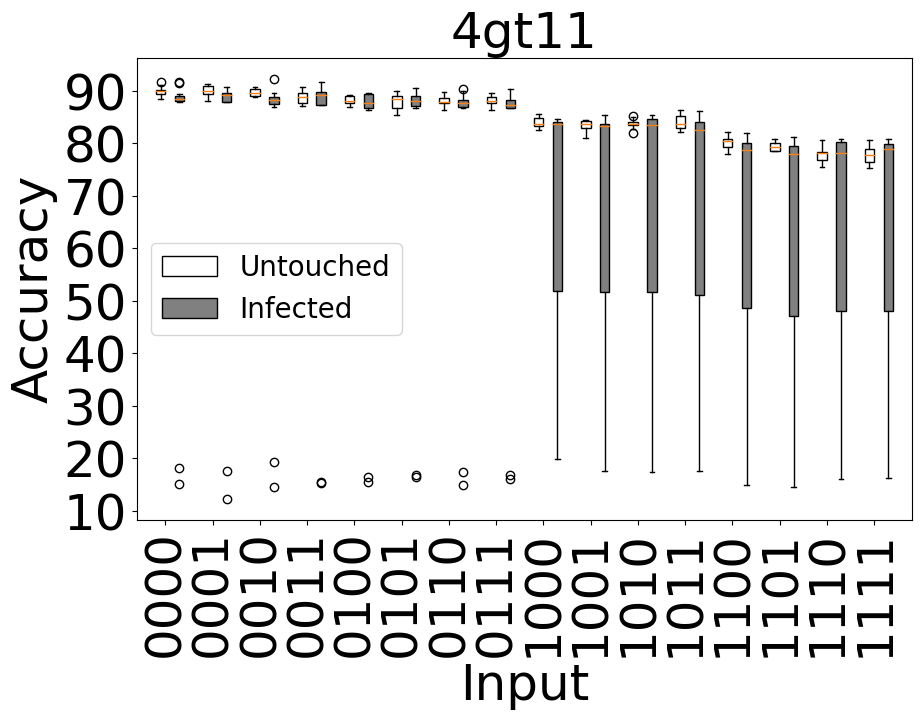}
                 \vspace{-4mm}
                 (e)
        \end{minipage}%
        \begin{minipage}{0.33\textwidth}
                \centering
                \includegraphics[width=\linewidth]{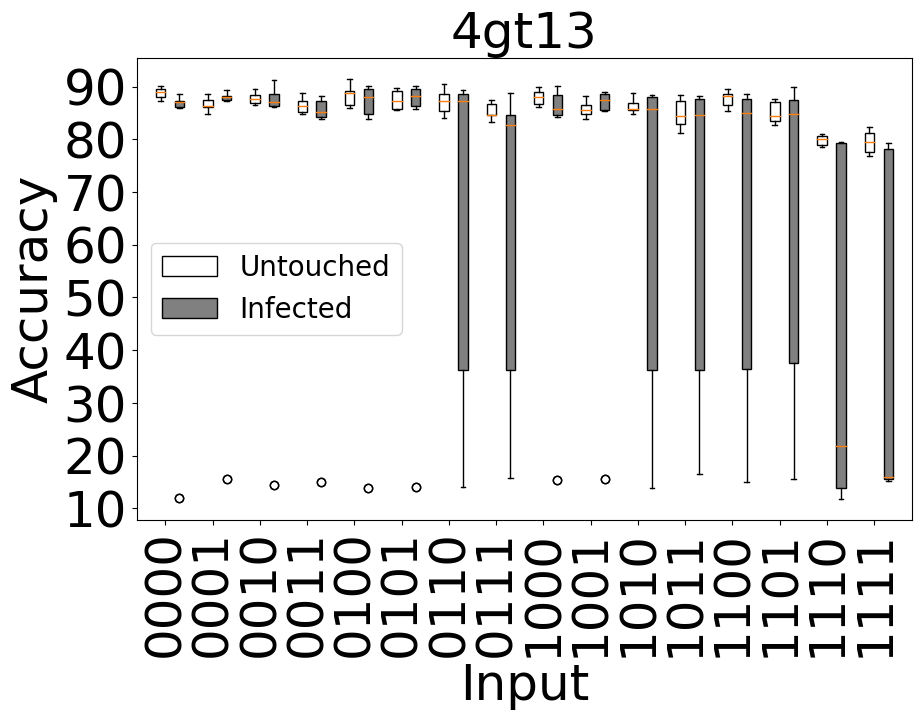}
                 \vspace{-4mm}
                 (f)
        \end{minipage}%
\end{figure*}
\section{Defense}
To defend quantum circuits from the proposed Trojan attack, we trained a modified form of a CNN based classifier to detect infected circuits and distinguish them from untouched benchmarks.
We discuss the basic design, modifications, and performance analysis of CNN based classifier for Trojan-infected circuits. We begin with the process of dataset generation and briefly talk about the architecture of the classifier. This is followed by the training and evaluation procedures used, and we then show the evaluation metrics and performance results of the model.
\subsection{Dataset Generation}
We first generated 50 benchmark circuits using Qiskit. The optimized circuits, called ‘untouched circuits’, were  compiled using Qiskit’s FakeValencia backend. We then insert the Trojan X gate at each possible location and generate 'Trojan Infected circuits'. They are also compiled in the same way. Thus, our dataset comprises 50 'untouched' and 313 'Trojan Infected' compiled circuits. We did the same for Hadamard infected quantum circuits.
\subsection{Architecture of Classifier}
The Classifier has been developed using the TensorFlow framework (the Keras API in particular) as a CNN model. 
%It it specifically designed to effectively distinguish between untouched and Trojan-infected compiled QAOA circuits by exploiting the intrinsic qualities of the circuit representations. 
It involves many key elements and methods to capture subtle features and patterns in the circuits. We leverage these qualities of the model to help distinguish our benchmark circuits from their infected versions. The CNN model is made of several layers that operate on the 2D representations of the quantum circuits created by transforming the Quantum Assembly Language (QASM) files for the circuits, to unitary matrices using Qiskit’s Operator method. These matrices denote the quantum gates in the circuit and are padded to a common size to ensure uniformity. 

The CNN consists of several layers. The first is a Convolutional layer for feature extraction, then a ReLU function to introduce non-linearity, followed by a MaxPooling layer to capture important features. The output of this layer is flattened and fed into a Dense layer to extract high-level features, to which ReLU is applied. The final output layer consists of two units with a softmax activation function, denoting the two classes: Untouched and Trojan infected. This function normalizes the outputs and provides probabilities for each class.
\subsection{Training and Evaluation}
The dataset is divided 95\%-5\% into a training-testing set for evaluating the model. The model's weights and biases are optimized iteratively. The modified model uses the Adam optimizer with learning rate scheduling, to help the model converge more steadily. We also apply L2 regularization to prevent overfitting, which adds a penalty term to the loss function, thus discouraging large weights. The CNN classifier learns to extract meaningful features and patterns from the circuit representations during training. The training dataset is split into validation and training sets, allowing the model to generalize well to unknown data.

%TrojanNet was designed for a balanced dataset, but 
In our use case, the data entered for the 2 output classes is not balanced since we have 50 'untouched' circuits and 313 'Trojan infected circuits'. This affects the training process and the model's performance, as the model becomes biased toward the majority class, in this case, the infected circuits.We used the technique of class weights to address this issue. By assigning different weights to each class during training, we give more importance to underrepresented classes and less importance to overrepresented classes. This helps the model pay more attention to the minority class and thus improve its ability to correctly classify examples from both classes. 
The model is evaluated using the validation dataset to supervise its performance and prevent overfitting during training. Various evaluation metrics are used to assess the performance of the model. Only measuring accuracy may not capture the model’s robustness in different scenarios. Hence, additional metrics, such as recall, precision and F1-score are considered, giving insights into the model’s performance in correctly distinguishing between untouched and Trojan-infected circuits. Precision refers to the proportion of correctly identified Trojan-inserted circuits among the predicted positives. Recall, also known as sensitivity, measures the proportion of Trojan-inserted circuits correctly identified among the true positives. The F1-score combines recall and precision, providing a balanced measure of the model’s performance.

\subsection{Results and Analysis}
The trained model is then evaluated using our generated datasets for the 2 types of Trojans being studied. Figs. \ref{accu_vs_epoch_not} and \ref{accu_vs_epoch_h} show the performance results in terms of accuracy and loss for both Trojans separately. The model was trained for 50 epochs. Training and validation accuracies gradually increased, reaching 94.7\% validation accuracy at epoch 23, and settling at around 90\% for the NOT infected circuits, while for the H gate infected circuits the accuracy reaches 89.47\% at epoch 13. Additional evaluation metrics such as precision, recall, and F1-score, were also calculated, amounting to 88.88\%, 100\% and 94.1\% respectively for the NOT infected circuits, and 83.33\%, 100\% and 90.9\% for the Hadamard infected circuits thus demonstrating that our modified version of TrojanNet works well for detection of Trojans in quantum circuits.

\section{Conclusions}
For efficient operation of Noisy Intermediate Scale Quantum (NISQ) computers, quantum circuits need to be optimized effectively using a compiler that decomposes high-level gates to basis gates. Several 3rd party compilers are emerging that are more efficient for complex quantum circuits, but they may be untrustworthy and pose security risks. 
%This could permit an adversary to apply a quantum Trojan attack during compilation that eludes detection through conventional tests, but shows up through its effects on the circuit output. 
In this paper, we study the impact of 2 single qubit Trojan gates(NOT \& Hadamard) in a benchmark suite from the Revlib library. We also presented a machine learning based detection of such Trojans(using the NOT infected circuits).

%In future works, we could use different multi-qubit gates, for Trojan attacks and characterise the impact from such Trojans. We could also use multiple single qubit gates as Trojans and observe how circuit behavior changes in response to such attacks.
\section{Acknowledgment}
This work is supported in parts by NSF (CNS-1722557, CNS-2129675, CCF-2210963, CCF-1718474, OIA-2040667, DGE-1723687, DGE-1821766 and DGE-2113839) and Intel’s gift.

\section*{}

\vspace{12pt}

\end{document}